\documentclass[twocolumn]{aastex63}

\usepackage{savesym}
\savesymbol{tablenum}
\usepackage{siunitx}
\restoresymbol{SIX}{tablenum}
\usepackage[bulgarian,english]{babel}
\usepackage[T2A]{fontenc}
\DeclareSIUnit\mEarth{M_\oplus}
\DeclareSIUnit\mSun{M_\odot}
\DeclareSIUnit\rEarth{R_\oplus}
\DeclareSIUnit\year{yr}
\DeclareSIUnit\au{au}
\DeclareSIUnit\dex{dex}
\DeclareSIUnit\ppm{ppm}

\newcommand{\starA}{TOI~4406}
\newcommand{\starB}{TOI~2338}
\newcommand{\starC}{TOI~2589}
\newcommand{\plA}{TOI~4406b}
\newcommand{\plB}{TOI~2338b}
\newcommand{\plC}{TOI~2589b}

\newcommand{\opA}{ \ensuremath{30.08364 \pm 0.00005\ {\rm d}} }
\newcommand{\opB}{ \ensuremath{22.65398 \pm 0.00002\ {\rm d}} }
\newcommand{\opC}{ \ensuremath{61.6277  \pm 0.0002\ {\rm d}} }

\newcommand{\mpA}{ \ensuremath{0.30 \pm 0.04\ \mjup} }
\newcommand{\mpB}{ \ensuremath{5.98 \pm 0.20\ \mjup} }
\newcommand{\mpC}{ \ensuremath{3.50 \pm 0.10\ \mjup} }

\newcommand{\rpA}{ \ensuremath{1.00 \pm 0.02\ \rjup} }
\newcommand{\rpB}{ \ensuremath{1.00 \pm 0.01\ \rjup} }
\newcommand{\rpC}{ \ensuremath{1.08 \pm 0.03\ \rjup} }

\newcommand{\epB}{ \ensuremath{0.676 \pm 0.002} }
\newcommand{\epC}{ \ensuremath{0.522 \pm 0.006} }

\newcommand{\feh}{\ensuremath{{\rm [Fe/H]}}}
\newcommand{\teff}{\ensuremath{T_{\rm eff}}}
\newcommand{\teq}{\ensuremath{T_{\rm eq}}}
\newcommand{\logg}{\ensuremath{\log{g}}}
\newcommand{\zaspe}{\texttt{ZASPE}}

\newcommand{\tess}{\textit{TESS}}
\newcommand{\vsini}{\ensuremath{v \sin{i}}}

\newcommand{\mjup}{\ensuremath{{\rm M_{J}}}}

\newcommand{\mpl}{\ensuremath{{\rm M_P}}}
\newcommand{\rjup}{\ensuremath{{\rm R_J}}}
\newcommand{\rpl}{\ensuremath{{\rm R_P}}}
\newcommand{\rstar}{\ensuremath{{\rm R}_{\star}}}
\newcommand{\mstar}{\ensuremath{{\rm M}_{\star}}}
\newcommand{\lstar}{\ensuremath{{\rm L}_{\star}}}
\newcommand{\rsun}{\ensuremath{{\rm R}_{\odot}}}
\newcommand{\msun}{\ensuremath{{\rm M}_{\odot}}}

\received{June 1, 2019}
\revised{January 10, 2019}
\accepted{\today}
\submitjournal{AJ}

\shorttitle{Three transiting warm giant planets}
\shortauthors{Brahm et al.}
\graphicspath{{./}}

\begin{document}

\title{Three long period transiting giant planets from \tess \footnote{Based on observations collected at the European Organization for Astronomical Research in the Southern Hemisphere under ESO programmes 0104.C-0413,	106.21ER.001, and 108.22A8.001, and MPG programmes 0103.A-9008, 0104.A-9007.}}

\correspondingauthor{Rafael Brahm}
\email{rafael.brahm@uai.cl}

\author[0000-0002-9158-7315]{Rafael Brahm}
\affil{Facultad de Ingeniera y Ciencias, Universidad Adolfo Ib\'{a}\~{n}ez, Av. Diagonal las Torres 2640, Pe\~{n}alol\'{e}n, Santiago, Chile}
\affil{Millennium Institute for Astrophysics, Chile}
\affil{Data Observatory Foundation, Chile}

\author[0000-0003-2417-7006]{Solène Ulmer-Moll}
\affiliation{Physikalisches Institut, University of Bern, Gesellsschaftstrasse 6, 3012 Bern, Switzerland}
\affiliation{Observatoire astronomique de l'Universit\'e de Gen\`eve, chemin Pegasi 51, 1290 Versoix, Switzerland}

\author[0000-0002-5945-7975]{Melissa J.\ Hobson} 
\affil{Max-Planck-Institut f\"{u}r Astronomie, K\"{o}nigstuhl  17, 69117 Heidelberg, Germany}
\affil{Millennium Institute for Astrophysics, Chile}

\author[0000-0002-5389-3944]{Andr\'es Jord\'an} 
\affil{Facultad de Ingeniera y Ciencias, Universidad Adolfo Ib\'{a}\~{n}ez, Av. Diagonal las Torres 2640, Pe\~{n}alol\'{e}n, Santiago, Chile}
\affil{Millennium Institute for Astrophysics, Chile}
\affil{Data Observatory Foundation, Chile}

\author[0000-0002-1493-300X]{Thomas Henning} 
\affil{Max-Planck-Institut f\"{u}r Astronomie, K\"{o}nigstuhl  17, 69117 Heidelberg, Germany}

\author[0000-0002-0236-775X]{Trifon Trifonov} 
\affil{Max-Planck-Institut f\"{u}r Astronomie, K\"{o}nigstuhl  17, 69117 Heidelberg, Germany}
\affiliation{Department of Astronomy, Sofia University ``St Kliment Ohridski'', 5 James Bourchier Blvd, BG-1164 Sofia, Bulgaria}

\author{Mat\'ias I. Jones} 
\affiliation{European Southern Observatory, Casilla 19001, Santiago, Chile}

\author[0000-0001-8355-2107]{Martin Schlecker} 
\affiliation{Steward Observatory and Department of Astronomy, The University of Arizona, Tucson, AZ 85721, USA}

\author[0000-0001-9513-1449]{Nestor Espinoza} 
\affil{Space Telescope Science Institute, 3700 San Martin Drive, Baltimore, MD 21218, USA}

\author[0000-0003-3047-6272]{Felipe I.\ Rojas} 
\affil{Instituto de Astrof\'isica, Facultad de F\'isica, Pontificia Universidad Cat\'olica de Chile, Chile}
\affil{Millennium Institute for Astrophysics, Chile}

\author[0000-0003-0974-210X]{Pascal Torres} 
\affil{Instituto de Astrof\'isica, Facultad de F\'isica, Pontificia Universidad Cat\'olica de Chile, Chile}
\affil{Millennium Institute for Astrophysics, Chile}

\author[0000-0001-8128-3126]{Paula Sarkis} 
\affil{Max-Planck-Institut f\"{u}r Astronomie, K\"{o}nigstuhl  17, 69117 Heidelberg, Germany}

\author{Marcelo Tala} 
\affil{Facultad de Ingeniera y Ciencias, Universidad Adolfo Ib\'{a}\~{n}ez, Av. Diagonal las Torres 2640, Pe\~{n}alol\'{e}n, Santiago, Chile}

\author[0000-0003-3130-2768]{Jan Eberhardt} 
\affil{Max-Planck-Institut f\"{u}r Astronomie, K\"{o}nigstuhl  17, 69117 Heidelberg, Germany}

\author[0000-0002-0436-7833]{Diana Kossakowski} 
\affil{Max-Planck-Institut f\"{u}r Astronomie, K\"{o}nigstuhl  17, 69117 Heidelberg, Germany}

\author[0000-0003-2186-234X]{Diego J. Muñoz} 
\affil{Facultad de Ingeniera y Ciencias, Universidad Adolfo Ib\'{a}\~{n}ez, Av. Diagonal las Torres 2640, Pe\~{n}alol\'{e}n, Santiago, Chile}
\affil{Center for Interdisciplinary Exploration and Research in Astrophysics (CIERA) and Department of Physics and Astronomy, Northwestern University, 2145 Sheridan Road, Evanston, IL 60208, USA}

\author[0000-0001-8732-6166]{Joel D. Hartman} 
\affiliation{Department of Astrophysical Sciences, Princeton University, 4 Ivy Lane, Princeton, NJ 08544}

\author{Gavin Boyle} 
\affiliation{El Sauce Observatory, Coquimbo Province, Chile}
\affil{Cavendish Laboratory, J J Thomson Avenue, Cambridge, CB3 0HE, UK}

\author[0000-0001-7070-3842]{Vincent Suc} 
\affiliation{Facultad de Ingeniera y Ciencias, Universidad Adolfo Ib\'{a}\~{n}ez, Av. Diagonal las Torres 2640, Pe\~{n}alol\'{e}n, Santiago, Chile}
\affil{El Sauce Observatory -- Obstech, Chile}

\author{Fran\c{c}ois Bouchy} 
\affiliation{Observatoire astronomique de l'Universit\'e de Gen\`eve, chemin Pegasi 51, 1290 Versoix, Switzerland}

\author{Adrien Deline} 
\affiliation{Observatoire astronomique de l'Universit\'e de Gen\`eve, chemin Pegasi 51, 1290 Versoix, Switzerland}

\author[0000-0003-4711-3099]{Guillaume Chaverot} 
\affiliation{Observatoire astronomique de l'Universit\'e de Gen\`eve, chemin Pegasi 51, 1290 Versoix, Switzerland}

\author{Nolan Grieves} 
\affiliation{Observatoire astronomique de l'Universit\'e de Gen\`eve, chemin Pegasi 51, 1290 Versoix, Switzerland}

\author{Monika Lendl} 
\affiliation{Observatoire astronomique de l'Universit\'e de Gen\`eve, chemin Pegasi 51, 1290 Versoix, Switzerland}

\author{Olga Suarez} 
\affiliation{Universit\'e C\^ote d'Azur, Observatoire de la C\^ote d'Azur, CNRS, Laboratoire Lagrange, Bd de l'Observatoire, CS 34229, 06304 Nice cedex 4, France}

\author[0000-0002-7188-8428]{Tristan Guillot} 
\affiliation{Universit\'e C\^ote d'Azur, Observatoire de la C\^ote d'Azur, CNRS, Laboratoire Lagrange, Bd de l'Observatoire, CS 34229, 06304 Nice cedex 4, France}

\author[0000-0002-5510-8751]{Amaury H.M.J. Triaud} 
\affiliation{School of Physics \& Astronomy, University of Birmingham, Edgbaston, Birmingham B15 2TT, United Kingdom}

\author{Nicolas Crouzet} 
\affiliation{European Space Agency (ESA), European Space Research and Technology Centre (ESTEC), Keplerlaan 1, 2201 AZ Noordwijk, The Netherlands}

\author[0000-0002-3937-630X]{Georgina Dransfield} 
\affiliation{School of Physics \& Astronomy, University of Birmingham, Edgbaston, Birmingham B15 2TT, United Kingdom}

\author[0000-0001-5383-9393]{Ryan Cloutier}
\affiliation{Dept. of Physics \& Astronomy, McMaster University, 1280 Main St West, Hamilton, ON, L8S 4L8, Canada}
\affiliation{Center for Astrophysics $\vert$ Harvard \& Smithsonian, 60 Garden St, Cambridge, MA, 02138, USA}

\author[0000-0003-1464-9276]{ Khalid Barkaoui} 

\affil{Astrobiology Research Unit, Universit\'e de Li\`ege, All\'ee du 6 Ao\^ut 19C, B-4000 Li\`ege, Belgium}
\affil{Department of Earth, Atmospheric and Planetary Science, Massachusetts Institute of Technology, 77 Massachusetts Avenue, Cambridge, MA 02139, USA} 
\affiliation{Instituto de Astrof\'isica de Canarias (IAC), 38200 La Laguna, Tenerife, Spain}

\author[0000-0001-8227-1020]{Rick P. Schwarz} 
\affiliation{Center for Astrophysics \textbar \ Harvard \& Smithsonian, 60 Garden St, Cambridge, MA 02138, USA}

\author[0000-0003-2163-1437]{Chris Stockdale} 
\affil{Hazelwood Observatory, Australia}

\author[0000-0002-3721-1683]{Mallory~Harris}
\affiliation{Department of Physics and Astronomy, University of New Mexico, 210 Yale Blvd NE, Albuquerque, NM 87106, USA}

\author[0000-0002-4510-2268]{Ismael~Mireles}
\affiliation{Department of Physics and Astronomy, University of New Mexico, 210 Yale Blvd NE, Albuquerque, NM 87106, USA}

\author[0000-0002-5674-2404]{Phil Evans} 
\affiliation{El Sauce Observatory, Coquimbo Province, Chile}
\author[0000-0003-3654-1602]{Andrew W. Mann} 
\affiliation{Department of Physics and Astronomy, The University of North Carolina at Chapel Hill, Chapel Hill, NC 27599, USA}

\author{Carl Ziegler} 
\affiliation{Department of Physics, Engineering and Astronomy, Stephen F. Austin State University, 1936 North Street, Nacogdoches, TX 75962}

\author{Diana Dragomir} 
\affil{Department of Physics and Astronomy, University of New Mexico, 1919 Lomas Blvd NE Albuquerque, NM, 87131, USA}

\author[0000-0001-6213-8804]{Steven Villanueva} 
\affiliation{NASA Goddard Space Flight Center, Exoplanets and Stellar Astrophysics Laboratory (Code 667), Greenbelt, MD 20771, USA}

\author{Christoph Mordasini} 
\affil{Center for Space and Habitability, Universitat Bern, Gesellshaftsstrasse 6, 3012 Bern, Switzerland}

\author{George Ricker} 
\affiliation{Department of Physics and Kavli Institute for Astrophysics and Space Research, Massachusetts Institute of Technology, Cambridge, MA 02139, USA}
\author[0000-0001-6763-6562]{Roland Vanderspek} 
\affiliation{Department of Physics and Kavli Institute for Astrophysics and Space Research, Massachusetts Institute of Technology, Cambridge, MA 02139, USA}
\author{David W. Latham}
\affiliation{Center for Astrophysics \textbar \ Harvard \& Smithsonian, 60 Garden St, Cambridge, MA 02138, USA}
\author[0000-0002-6892-6948]{Sara Seager} 
\affiliation{Department of Physics and Kavli Institute for Astrophysics and Space Research, Massachusetts Institute of Technology, Cambridge, MA 02139, USA}
\affiliation{Department of Earth, Atmospheric and Planetary Sciences, Massachusetts Institute of Technology, Cambridge, MA 02139, USA}
\affiliation{Department of Aeronautics and Astronautics, MIT, 77 Massachusetts Avenue, Cambridge, MA 02139, USA}
\author{Joshua N. Winn} 
\affiliation{Department of Astrophysical Sciences, Princeton University, 4 Ivy Lane, Princeton, NJ 08544}

\author{Jon M. Jenkins} 
\affiliation{NASA Ames Research Center, Moffett Field, CA 94035, USA}

\author{Michael~Vezie} 
\affiliation{Department of Physics and Kavli Institute for Astrophysics and Space Research, Massachusetts Institute of Technology, Cambridge, MA 02139, USA}

\author{Allison Youngblood} 
\affiliation{NASA Goddard Space Flight Center, 8800 Greenbelt Rd, Greenbelt, MD 20771, USA}

\author[0000-0002-6939-9211]{Tansu~Daylan} 
\affiliation{Department of Astrophysical Sciences, Princeton University, 4 Ivy Lane, Princeton, NJ 08544}
\affiliation{LSSTC Catalyst Fellow}

\author[0000-0001-6588-9574]{Karen~A.~Collins} 
\affiliation{Center for Astrophysics \textbar \ Harvard \& Smithsonian, 60 Garden St, Cambridge, MA 02138, USA}

\author[0000-0003-1963-9616]{Douglas A. Caldwell} 
\affiliation{SETI Institute/NASA Ames Research Center}

\author[0000-0002-5741-3047]{David~ R.~Ciardi} 
\affiliation{Caltech/IPAC-NASA Exoplanet Science Institute, 770 S. Wilson Avenue, Pasadena, CA 91106, USA}

\author[0000-0003-0987-1593]{Enric Palle} 
\affiliation{Instituto de Astrof\'isica de Canarias (IAC), 38200 La Laguna, Tenerife, Spain}

\author[0000-0001-9087-1245]{Felipe Murgas} 
\affiliation{Instituto de Astrof\'isica de Canarias (IAC), 38200 La Laguna, Tenerife, Spain}
\affiliation{Departamento de Astrof\'isica, Universidad de La Laguna (ULL), E-38206 La Laguna, Tenerife, Spain}



\begin{abstract}

We report the discovery and orbital characterization of three new transiting warm giant planets. These systems were initially identified as presenting single transit events in the light curves generated from the full frame images of the Transiting Exoplanet Survey Satellite (\tess). Follow-up radial velocity measurements and additional light curves were used to determine the orbital periods and confirm the planetary nature of the candidates. The planets orbit slightly metal-rich late F- and early G-type stars. We find that \plA\ has a mass of \mpl=\mpA , a radius of \rpl=\rpA, and a low eccentricity orbit ($e$=0.15$\pm$0.05) with a period of P$=$\opA. \plB\ has a mass of \mpl=\mpB, a radius of \rpl=\rpB, and a highly eccentric orbit ($e$=\epB) with a period of P=\opB. Finally, \plC\ has a mass of \mpl=\mpC, a radius of \rpl=\rpC, and an eccentric orbit ($e$ =\epC) with a period of P=\opC. \plA\ and \plB\ are enriched in metals compared to their host stars, while the structure of \plC\ is consistent with having similar metal enrichment to its host star. 
\end{abstract}

\keywords{}


\section{Introduction} \label{sec:intro}
In the past couple of decades astronomers have started to uncover the different types of planets that orbit stars other than the Sun, finding that the solar system architecture is just one of the many possible outcomes of the planet formation process \citep[see][for a detailed review]{winn:2015}. Among the new types of planets identified, gas giant planets with short orbits \citep[e.g.][]{mayor:1995,bakos:2007} stand out because they were not expected to exist according to classical models of planet formation \citep[e.g.][]{stevenson:1982}. While advances have been made in observational and theoretical aspects, the formation mechanism of this planet population is still an unsolved puzzle.

If these planets were not formed at their current locations, they should have migrated from beyond the snow-line. One possible migration scenario involves the loss of angular momentum through torques with the gaseous disc as long as it is still present \citep[e.g.][]{goldreich:1980,lin:1986}. Another possible migration mechanism is through gravitational interactions with a third body that excites the orbital eccentricity, which generates orbital shrinking due to tidal interactions with the star during periastron passages \citep[e.g.][]{rasio:1996,wu:2011}.

These two types of migration processes predict different outcomes in terms of orbital parameters for the inner planet. Specifically, ongoing high eccentricity migration mechanisms predict a wide distribution for the orbital eccentricities and obliquities \citep[e.g.][]{chatterjee:2008}, while disc migration by itself mostly predicts circular orbits aligned with the stellar spin. Additionally, the bulk structures and atmospheric compositions of these planets can be related to formation locations in the disc and possible migration paths \citep{Madhusudhan:2014,molliere:2022}. The detection and detailed characterization of the physical and orbital properties of close-in giant planets is crucial for constraining their formation and migration scenarios. In this regard, those giant planets that transit bright stars are particularly valuable because of the amount of information that can be obtained. 

Hot Jupiters, broadly defined as giant planets with orbital periods shorter than 10 days, have been efficiently characterized in the past decades \citep[see][for a detailed review]{dawson:2018}, allowing one to get some constraints on the main migration mechanisms involved in their orbital evolution \citep[e.g.][]{rice:2022}. Nonetheless, due to their extreme proximity to the star, conclusions based on this population can be affected by posterior radiative and tidal interactions with the star \citep[e.g.][]{albrecht:2012}. On the other hand, giant planets with orbital periods longer than 10 days but inside the snow-line are not subject to strong proximity effects, which allows one to perform a more direct comparison of their properties to those predicted by the different formation/migration scenarios \citep[e.g.][]{huang:2016}. For example, the mass and radius of warm Jupiters can be used to infer the bulk internal structure and composition of the planet through the use of standard interior models \citep{espinoza:2017}. Additionally the obliquity angle measured for warm Jupiters through the Rossiter-McLaughlin effect can be used to constrain migration models \citep{morton:2011}.

While ground-based photometric surveys have been successful in the detection of transiting hot Jupiters around bright stars, those having periods longer than $\sim$8 days, with some particular exceptions \citep[e.g. HATS-17b, ][]{brahm:2016}, remained mostly elusive to these projects due to the diurnal cycle.
The NASA Kepler mission started to expand the parameter space of well characterized giant planets by discovering some candidates orbiting stars bright enough for radial velocity measurements to determine their masses \citep{almenara:2015,dalba:2021,dalba:2021b,chachan:2022}. 
Nonetheless, due to its increased field of view compared to that of the NASA Kepler mission, the Transiting Exoplanet Survey Satellite \citep[\tess,][]{ricker:2015}, is enabling for the first time the systematic exploration and characterization of the domain of transiting warm Jupiters, either through the detection of periodic transiting signals or through the detection of single transiters \citep[e.g.][]{gill:2020,dawson:2021,dong:2021,eisner:2021,dalba:2022,grieves:2022}. The brightness of the stars monitored by \tess\ allows a detailed dynamical characterization to be obtained via the measurement of precision radial velocities. In this context, the Warm gIaNts with \tess\ (WINE) collaboration is currently performing a dedicated survey to identify, confirm and characterize warm giant planets using as a starting point the \tess\ data \citep{brahm:2019,jordan:2020,brahm:2020,schlecker:2020,hobson:2021,trifonov:2021}.

In this study, we present the discovery, confirmation, and orbital characterization of three new transiting warm giant planets initially identified as community \tess\ objects of interest (cTOIs) by the WINE collaboration and subsequently adopted by the \tess\ project as project TOIs. We have utilized ground based spectroscopic and photometric observations to confirm these planetary candidates as three new warm giant planets.

\section{Observations} \label{sec:obs}

\subsection{\tess}
All three transiting candidates presented in this study were identified from the Full Frame Image (FFI) light curves of the \tess\ primary mission.
The FFIs were calibrated by the Science Processing Operations Center (SPOC) at NASA Ames Research Center \citep{jenkins:2016}.
As the WINE collaboration, we generate light curves through the \texttt{tesseract\footnote{https://github.com/astrofelipe/tesseract}} pipeline for all stars brighter than T$=$13 mag of all \tess\ sectors soon after the data is made public. Due to our focus on the detection of giant transiting giant planets in long period orbits (P$>$10 d) we process all the light curves with a dedicated algorithm that after passing them through a median filter, identifies systematic negative deviations in flux, whose amplitudes could be consistent with the depths and durations of transits of giant planets orbiting  main sequence/subgiant stars.  All systems that present such specific variations are then manually vetted. The light curves from which the long period candidates are identified are not corrected for flux contamination of neighbouring stars by \texttt{tesseract}. The possible dilution of the transits by neighbouring stars that fall inside the \tess\ photometric aperture is considered in the manual vetting process when estimating the planet radius from the transit depth and host stellar radius.

\starA\ was observed by \tess\ in sector 2. An $\approx$5100 ppm single transit was identified with a duration of $\approx$0.16 d. A query to the GAIA DR2 \citep{gaia:dr2} archive revealed that this target presented no neighbouring stars closer than 30", and a small uncertainty in the radial velocity measurements (0.6 km s$^{-1}$), and was therefore selected as a high priority candidate of the WINE collaboration, suitable for spectroscopic follow-up.

\starB\ was observed by \tess\ in sectors 5 and 6. A single transiter candidate was identified in the \texttt{tesseract} light curve of sector 5. The transit-like signal presented a depth of $\approx$11000 ppm, and a duration of 0.17 days. As soon as the FFIs of sector 6 were made public, we generated the corresponding light curve for \starB\ and identified a second transit with similar properties to those of sector 5. The transits were separated by 22.7 days. GAIA DR2 reports just one faint neighbouring star closer than 30" for \starB, and a small radial velocity uncertainty (0.6 km s$^{-1}$), and therefore this system was also classified as a high priority WINE candidate.

 \starC\ was observed by \tess\ in sector 7. In its \texttt{tesseract} light curve we identified a single transiter with a depth of $\approx 9000$ ppm, and a duration of 0.28 days. As opposed to the other two candidates, \starC\ presented 9 neighbouring stars closer than 30" according to the GAIA DR2 catalog. Nonetheless all of them were relatively faint in comparison to \starC. The reported radial velocity uncertainty was small (0.6 km s$^{-1}$) and this system was also selected as a high priority candidate. 

Due to their promising properties we proposed these three systems (along with others) to be observed in the Cycle 3 of the \tess\ mission in 2-minute cadence mode in order to further constrain the transiting parameters and determine the orbital periods of the candidates orbiting \starA\ and \starC. Therefore, in the first year of the \tess\ extended mission, additional \tess\ light curves were obtained for these three systems. \starA\ was observed in 2-minute cadence mode in sectors 28 and 29, presenting one additional transit in each light curve. \starB\ was observed in sector 32 in 2-minute cadence  mode, and one additional transit was identified for this candidate. Finally, \starC\ was observed by \tess\ in 2-minute mode in sectors 33 and 34, and an additional transit was identified in sector 34. These new \tess\ data unambiguously constrained the periods of  \starA\ and \starB\ to be of 30.08 d and 22.65 d each, but were not enough to determine a exact orbital period for \starC, for which additional data was required.

All these 2-minute cadence observations were processed by the SPOC pipeline \citep{twicken:2018}.
For \starA\ the SPOC search of Sectors 28 and 29 generated a clean Data Validation report and the transit signature passed all the diagnostic tests, including the difference imaging centroiding test, which located the source of the transits to within 1.233+/-2.6804 arcsec of the target star’s catalog position.
For \starB\ the SPOC search of sector 34 identified the single transit and the difference image centroiding test located the source of the transit to within 0.3 +/- 2.5 arcsec of the catalog star position.
 Finally, for \starC\ the SPOC search of sector 32 identified the single transit and the difference image centroding test located the source of the transit to within 0.35 +- 2.7 arcsec of the catalog star position.

\begin{figure*}[tp]
    \centering
    \includegraphics[width=\textwidth]{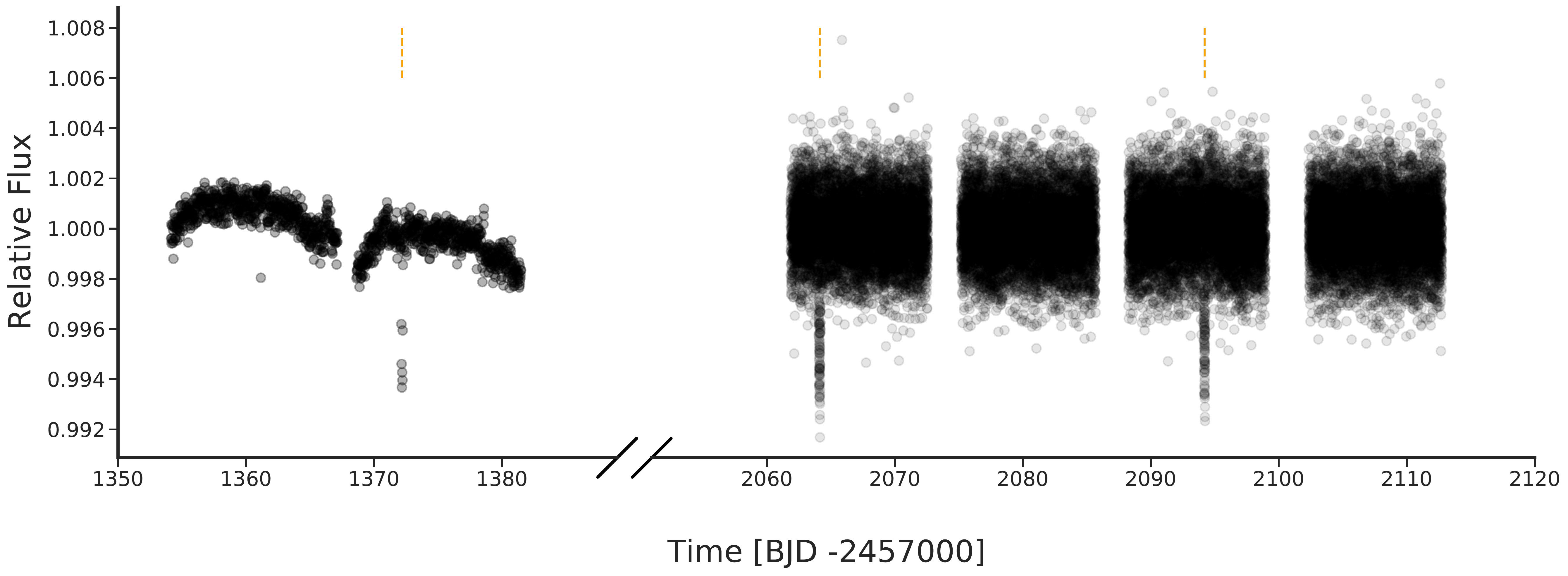}
    \caption{\tess\ light curves of \starA. Dashed vertical lines indicate the transits that were identified.}
    \label{tessA} 
\end{figure*}

\begin{figure*}[tp]
    \centering
    \includegraphics[width=\textwidth]{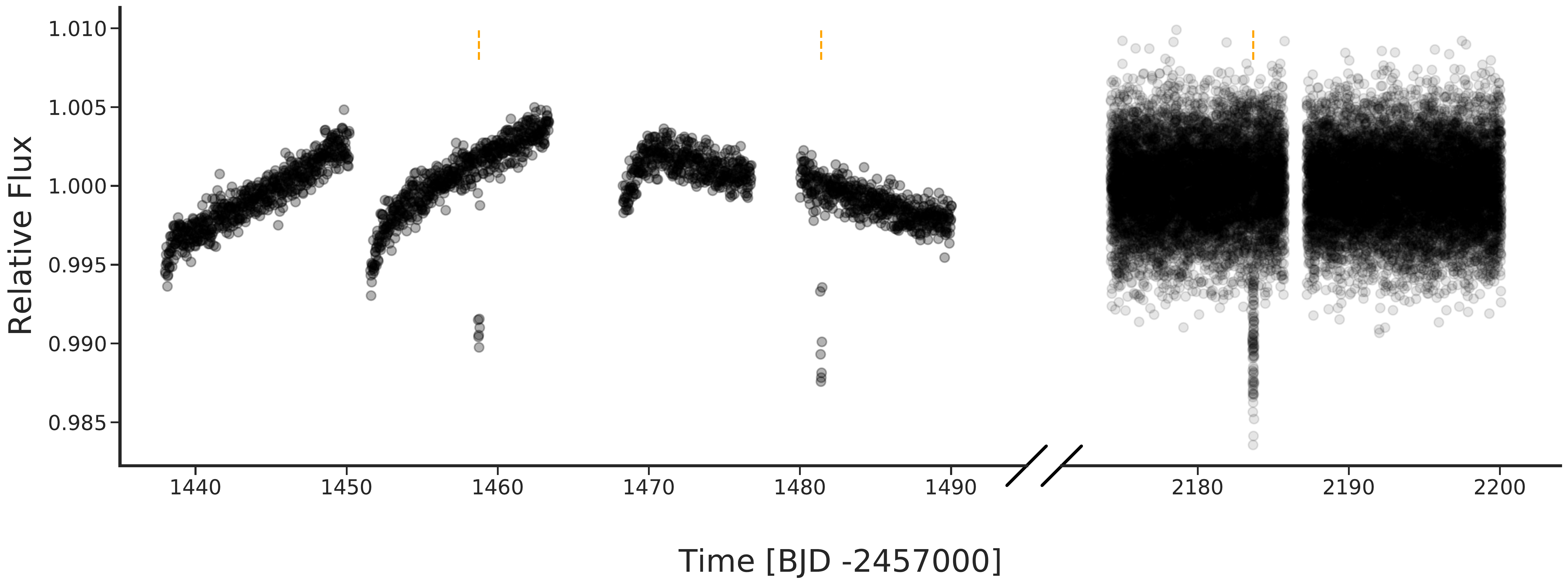}
    \caption{\tess\ light curves of \starB. Dashed vertical lines indicate the transits that were identified.}
    \label{tessB} 
\end{figure*}

\begin{figure*}[tp]
    \centering
    \includegraphics[width=\textwidth]{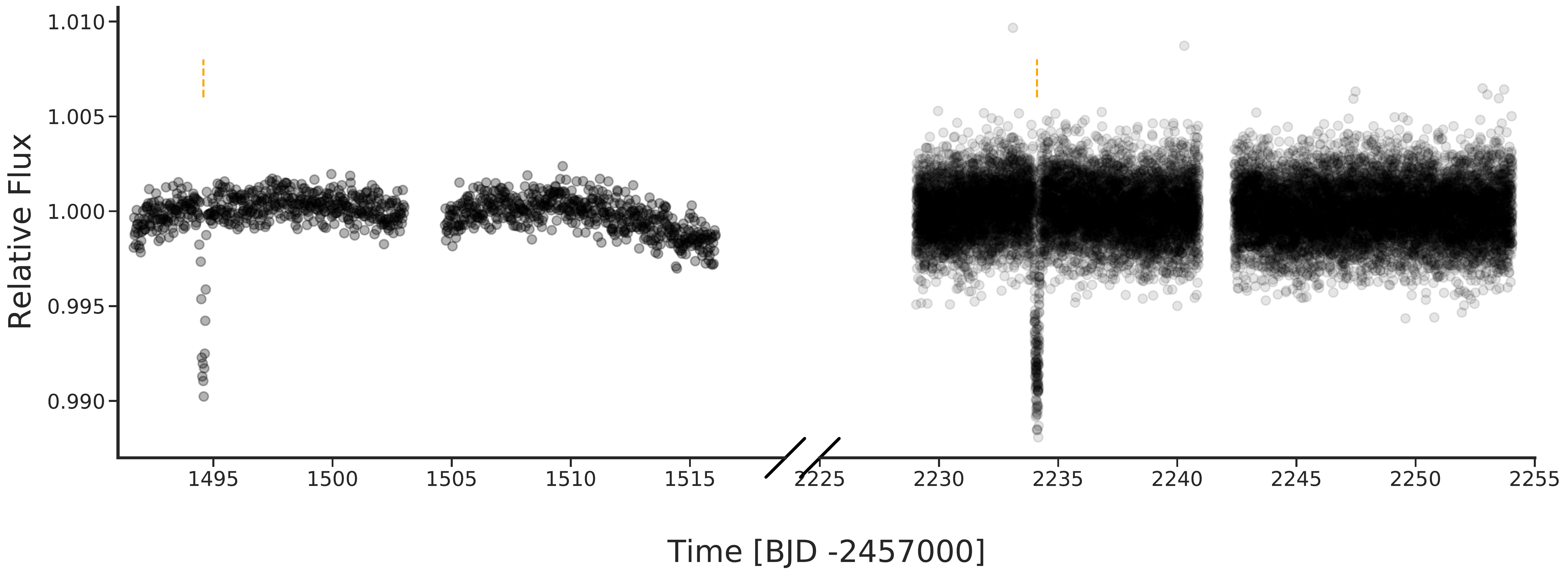}
    \caption{\tess\ light curves of \starC. Dashed vertical lines indicate the transits that were identified.}
    \label{tessC} 
\end{figure*}

Figures \ref{tessA}, \ref{tessB}, and \ref{tessC}
show all \tess\ light curves with transits for \starA, \starB, and \starC, respectively.

\subsection{Spectroscopy}
\begin{figure}[tp]
    \centering
    \includegraphics[width=\columnwidth]{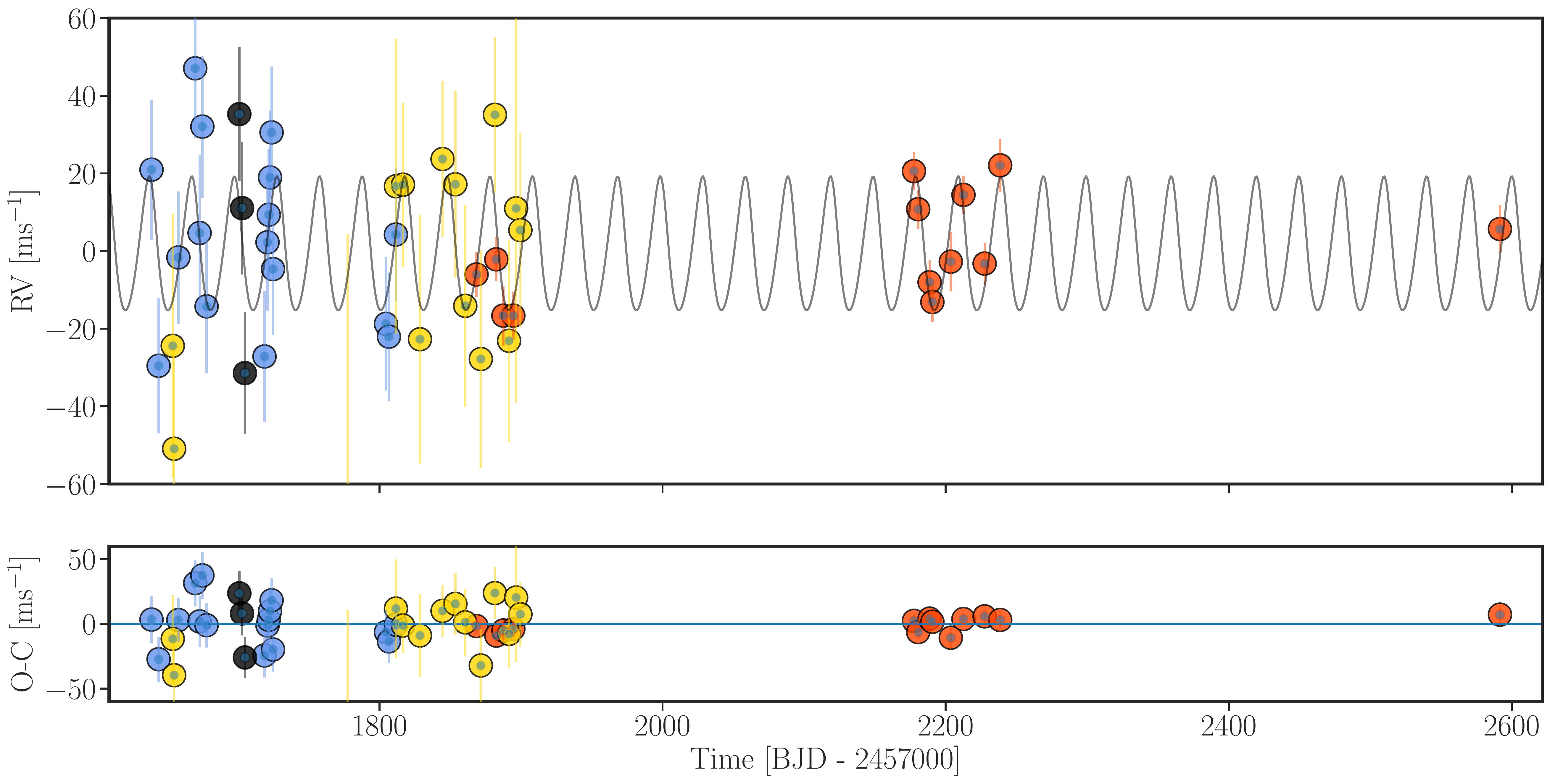}
    \caption{Radial velocities of \starA\ obtained with FEROS (blue), Coralie (yellow), CHIRON (black), and HARPS (orange).}
    \label{starA-rvs} 
\end{figure}

\begin{figure}[tp]
    \centering
    \includegraphics[width=\columnwidth]{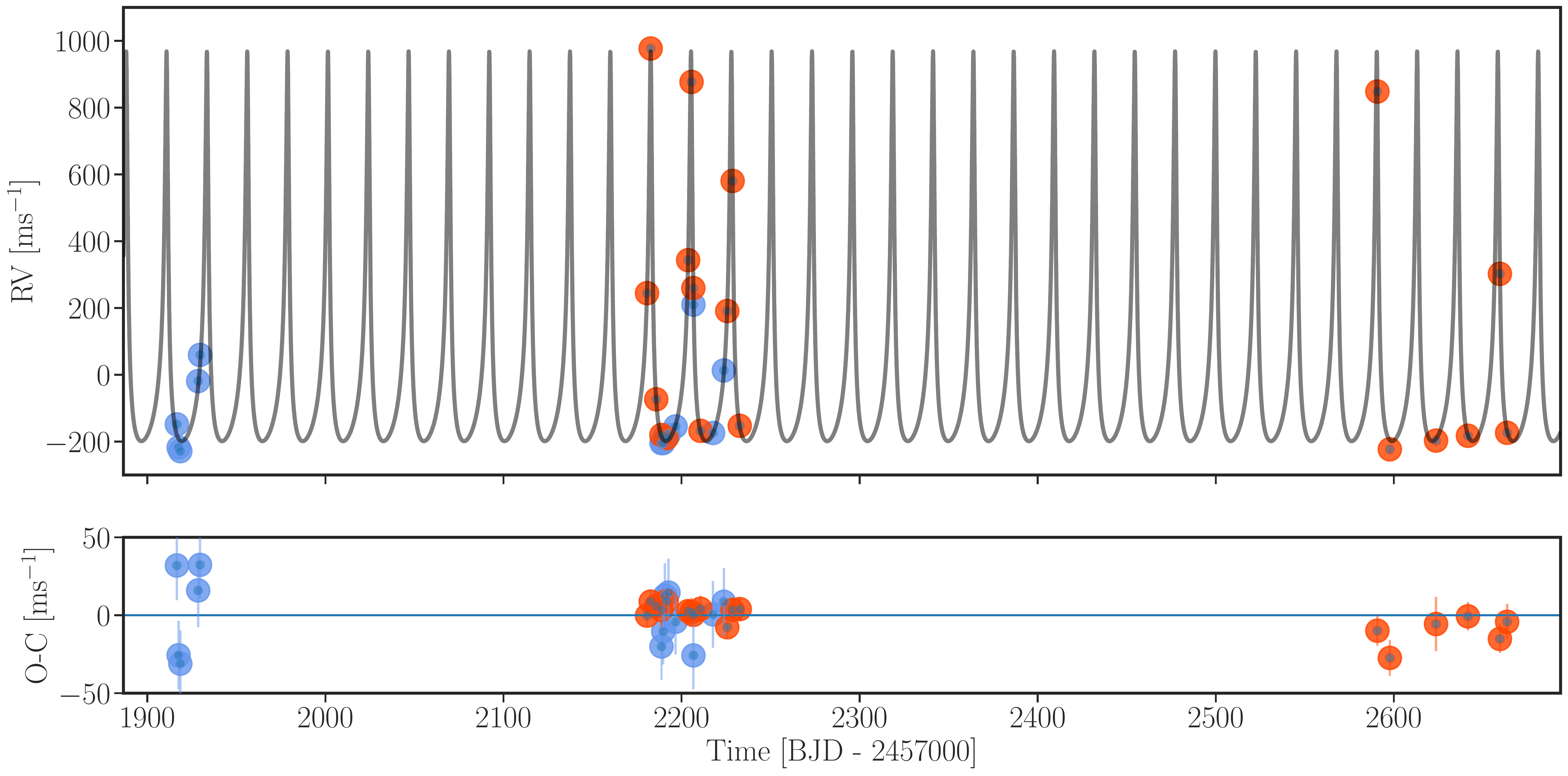}
    \caption{Radial velocities of \starB\ obtained with FEROS (blue) and HARPS (orange).}
    \label{starB-rvs} 
\end{figure}

\begin{figure}[tp]
    \centering
    \includegraphics[width=\columnwidth]{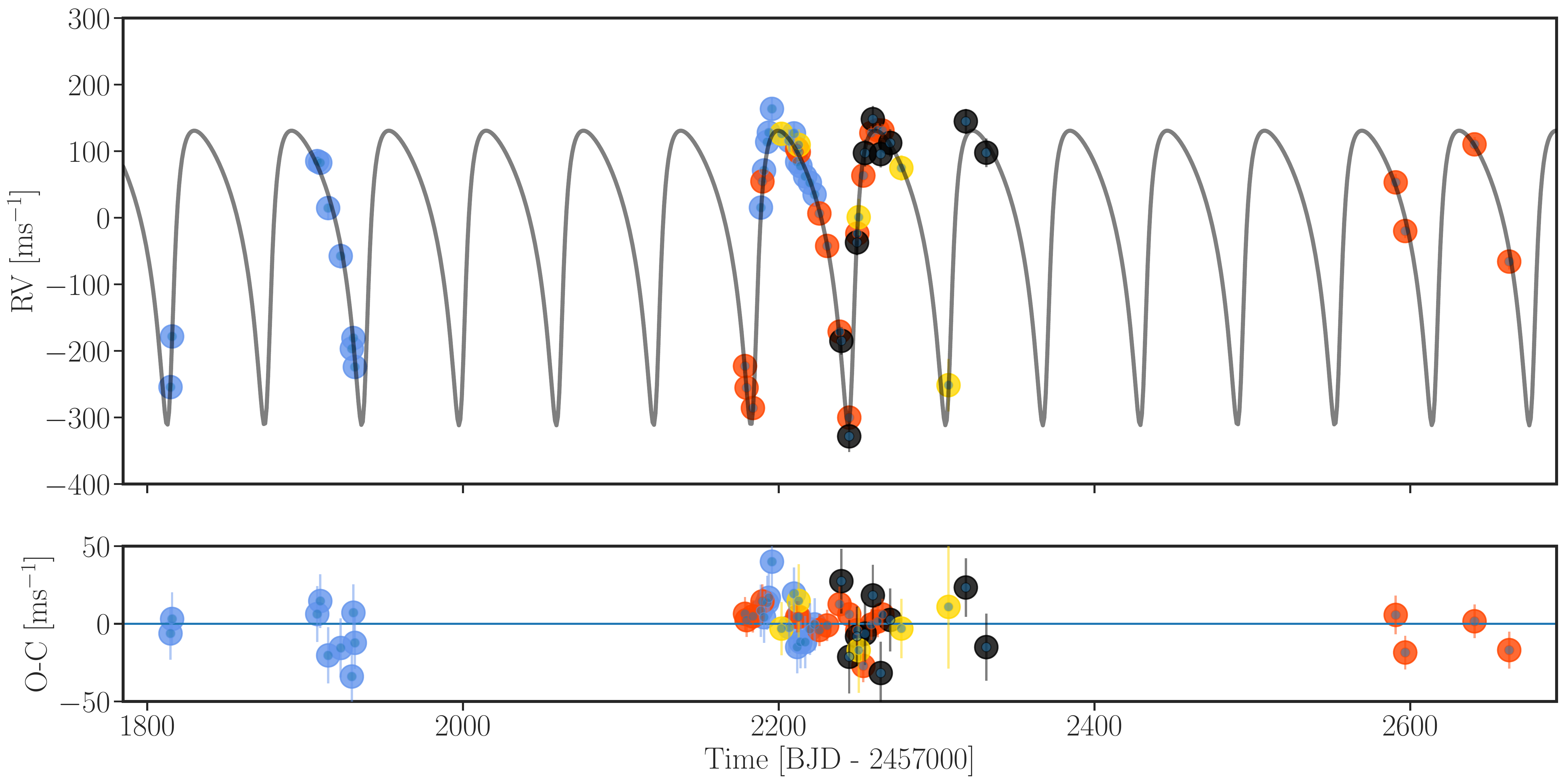}
    \caption{Radial velocities of \starC\ obtained with FEROS (blue), CHIRON (black), HARPS (orange), and CORALIE (yellow).}
    \label{starC-rvs} 
\end{figure}

Our three targets of interest were followed up spectroscopically with four different instruments installed in the north of Chile. The goal of these follow-up observations was to measure precision radial velocities and search for variations consistent with the planetary hypothesis for the transiting signals. All the radial velocities obtained for our three systems are presented in table \ref{tab:rvs}. The radial velocity measurements are also presented in Figures \ref{starA-rvs}, \ref{starB-rvs}, and \ref{starC-rvs}, for \starA, \starB, and \starC, respectively.

\begin{deluxetable*}{lrrrrrrr}
\tablewidth{0pc}
\tablecaption{
    Radial velocity measurements for \starA, \starB, and \starC. \label{tab:rvs}
}
\tablehead{
    \colhead{ID} &
    \colhead{BJD} & 
    \colhead{RV} & 
    \colhead{\ensuremath{\sigma_{\rm RV}}} &
    \colhead{BIS} & 
    \colhead{\ensuremath{\sigma_{\rm BIS}}} &
    \colhead{$S_{index}$} & 
    \colhead{Instrument} \\
    \colhead{} &    
    \colhead{-2450000} & 
    \colhead{(m s$^{-1}$)} & 
    \colhead{(m s$^{-1}$)} &
    \colhead{(m s$^{-1}$)} & 
    \colhead{(m s$^{-1}$)} &
    \colhead{} & 
    \colhead{} \\    
}
\startdata
\plA & 8638.90562 & 24211.8 & 10.5 & 1.0 & 12.0 & 0.138 $\pm$ 0.013 & FEROS \\
\plA & 8643.92198 & 24161.3 & 9.4 & 32.0 & 11.0 & 0.173 $\pm$ 0.011 & FEROS \\
\plA & 8657.87399 & 24189.2 & 8.6 & -9.0 & 10.0 & 0.140 $\pm$ 0.009 & FEROS \\
\plA & 8669.78376 & 24237.9 & 10.3 & 18.0 & 11.0 & 0.135 $\pm$ 0.012 & FEROS \\
\plA & 8672.86255 & 24195.5 & 13.6 & -6.0 & 14.0 & 0.151 $\pm$ 0.018 & FEROS \\ 
\plA & 8674.83206 & 24222.9 & 10.9 & -1.0 & 12.0 & 0.107 $\pm$ 0.012 & FEROS \\
\plA & 8677.80976 & 24176.6 & 9.0 &   3.0 & 10.0 & 0.139 $\pm$ 0.009 & FEROS \\
\plA & 8718.74463 & 24163.7 & 8.4 &  17.0 & 10.0 & 0.142 $\pm$ 0.008 & FEROS \\
\plC & 8814.78468 & 68294.7 & 6.6 & -19.0 & 11.0 & 0.146 $\pm$ 0.010 & FEROS \\
\plC & 8815.78744 & 68370.8 & 6.9 & -32.0 & 11.0 & 0.182 $\pm$ 0.011 & FEROS \\
\plC & 8907.71907 & 68634.2 & 8.8 &  26.0 & 13.0 & 0.184 $\pm$ 0.020 & FEROS \\
\plC & 8909.64069 & 68631.8 & 7.5 &  -9.0 & 11.0 & 0.139 $\pm$ 0.012 & FEROS \\
\plC & 8914.63041 & 68563.7 & 9.1 & -18.0 & 13.0 & 0.165 $\pm$ 0.017 & FEROS \\
\plC & 8922.59456 & 68491.6 & 10.9 &  9.0 & 15.0 & 0.176 $\pm$ 0.023 & FEROS \\
\plB & 8916.59777 &	83777.2 & 10.3 & -31.0 & 15.0 & 0.229 $\pm$ 0.023 & FEROS \\
\plB & 8917.57140 &	83708.0 & 9.8  & -27.0 & 14.0 & 0.219 $\pm$ 0.023 & FEROS \\
\plB & 8918.53080 &	83697.1 & 8.2  &   1.0 & 12.0 & 0.163 $\pm$ 0.016 & FEROS \\
\plB & 8928.56100 &	83906.9 & 12.9 &   1.0 & 18.0 & 0.154 $\pm$ 0.041 & FEROS \\
\plB & 8929.57890 &	83985.2 & 8.2  &   5.0 & 12.0 & 0.173 $\pm$ 0.017 & FEROS \\
\plB & 9180.61110 &	84186.1 & 5.5  & 15.0 & 7.0 & 0.033 $\pm$ 0.013 & HARPS \\
\plB & 9182.66670 &	84918.6 & 6.7  &  5.0 & 9.0 & -0.008 $\pm$ 0.014 & HARPS \\
\enddata
\tablecomments{
    This table is available in a machine-readable form in the online
    journal.  A portion is shown here for guidance regarding its form
    and content.
}
\end{deluxetable*}

\subsubsection{FEROS}
The Fibre-fed, Extended Range, Échelle Spectrograph \citep[FEROS,][]{kaufer:99} is a bench-mounted, thermally controlled, prism-cross-dispersed échelle spectrograph, currently installed in the MPG 2.2m telescope at the ESO La Silla Observatory, in Chile. It has a spectral resolution of R$\approx$48000, high efficiency (20\%), and a spectral coverage from 350 nm to 920 nm divided in 39 echelle orders. FEROS was the first instrument to be used in the spectroscopic follow-up of the three systems presented in this study. All FEROS observations associated to the WINE collaboration are performed with the simultaneous calibration mode \citep{baranne:96} for tracing instrumental velocity drift variations in the spectrograph enclosure due to changes in its environment (mostly temperature and atmospheric pressure changes).


\starA\ was observed on 16 different epochs between June and November 2019. The adopted exposure time was of 600s which delivered spectra with signal-to-noise ratios per resolution element in the range of 50 to 100 depending on the specific observing conditions of each night.

\starB\ was observed on 13 different nights with FEROS starting in March of 2020, and taking the last spectrum in January of 2021. The exposure time was set to 1200s for this target, and we obtained spectra with signal-to-noise ratios between 50 and 80.

We obtained 21 FEROS spectra of \starC\ between November of 2019 and January of 2021. In this case we adopted an exposure time of 900s, which translated to a typical signal-to-noise ratio of 80.

All FEROS data obtained for this study were processed with the \texttt{ceres} pipeline \citep{brahm:2017:ceres}. \texttt{ceres} performs all the reduction steps to generate a continuum normalized, wavelength calibrated, and optimally extracted spectrum from raw data. \texttt{ceres} also computes precision radial velocities and bisector span measurements through the cross-correlation technique. A G2-type binary mask was used as a template for computing the cross-corelation function of the spectra of our three candidates.

FEROS radial velocities for \starA\ indicated that some time-correlated variations were present but with an amplitude not much larger than the radial velocity uncertainties. For \starB, FEROS radial velocities quickly showed significant variations in phase with the P=22 days orbital period of the transiting candidate, with an amplitude that could be produced by a giant planet. Finally, FEROS radial velocities for \starC\ also showed variations that could be produced by a massive planet, and after some months of observations, we were able to constrain the orbital period of this single transit candidate from these radial velocities.

\subsubsection{HARPS}
The High Accuracy Radial velocity Planet Searcher \citep[HARPS,][]{mayor:2003} is a high resolution and stabilized spectrograph fibre-fed by the Cassegrain focus of the ESO\~3.6m telescope of the ESO La Silla Observatory, in Chile. It covers the spectral region of 380 nm to 690 nm, with a resolving power of R=115,000. We decided to use HARPS to follow-up our three systems due to different reasons. The amplitude of the tentative radial velocity variations that we identified with FEROS for \starA\ was too shallow to tightly constrain a possible orbit. In the case of \starB, even though the amplitude of the radial velocity signal was significantly larger, we did not have access to FEROS during the orbital phase close to periastron, so that the eccentricity was significantly unconstrained with the FEROS data alone. Finally, for \starC, we initially identified a second small-amplitude variation in addition to the one linked to the transiting planet, and we were interested to see if we could constrain it with HARPS radial velocities.
All HARPS data for this work were processed with the \texttt{ceres} pipeline using the same binary mask that was used for the processing of FEROS data.

We obtained thirteen HARPS spectra of \starA\ between January of 2020 and January of 2022. We adopted an exposure time of 1200 s for these observations, which yielded spectra with a signal-to-noise ratio of 55.
\starB\ was observed on eighteen different epochs, between November of 2020 and April of 2022. The adopted exposure time for this object was of 1200 s, which resulted in a typical signal-to-noise ratio of 30.
Finally, \starC\ was observed nineteen times with HARPS between November of 2020 and April of 2022. The resulting spectra presented signal-to-noise ratios around 45, which was obtained with an exposure time of 900 s.

\subsubsection{CHIRON}
CHIRON is a high resolution spectrograph installed in the 1.5 m telescope at the Cerro Tololo International Observatory.
We obtained three spectra of \starA\ with the CHIRON high-resolution spectrograph \citep{chiron}, in August of 2019. The data were obtained with the image slicer (R $\sim$\,80000), and an exposure time of 1800 s, leading to a SNR per pixel of $\sim$\,20 at 550 nm. Similarly, we collected a total of 15 spectra of \starC\ between January 24th, 2021 and January 18th, 2022. The adopted exposure time was also 1800 s. In the two cases we obtained a ThAr spectrum immediately before the science spectra to account for the instrument night drift. The data were reduced by the CHIRON pipeline \citep{Paredes2021} and the radial velocities were computed using an updated version of the pipeline used in \citet{Jones2019}.

\subsubsection{CORALIE}
The Swiss 1.2m Leonhard Euler Telescope installed at the ESO La Silla Observatory hosts the high resolution CORALIE spectrograph \citep{coralie:1998}. CORALIE works at a resolution of R=60000 and uses the simultaneous calibration technique by allowing the science exposures to count with a second fibre illuminated by the spectrum of a Thorium Argon lamp or by a Fabry-Perot system. We obtained 18 CORALIE spectra of \starA\ using an exposure time of 1200 s, and 5 spectra of \starC\ with an exposure time of 1800 s. The selected comparison source was the Fabry-Perot system, and the data was processed with the standard Data Reduction Software (DRS) of the instrument.

\subsection{Ground-based photometry}
Ground-based photometric time series obtained with sub-1m telescopes are commonly used to confirm the planetary nature of transiting candidates identified by \tess. These light curves are used to confirm that the transit-like feature occurs on the particular star of interest, and not on another star inside the photometric aperture of the \tess\ data. Given the low number of transits present in \tess\ data for the three systems presented in this study, these additional light curves are also crucial for refining the photometric ephemeris. The observations were obtained in the context of the \tess\ follow-up program (TFOP) subgroup 1.

We used the Las Cumbres Observatory \citep[LCO,][]{brown:2013} Network of telescopes to obtain follow-up light curves of \starA\ and \starB. For this goal we used the LCO stations located in the South African Astronomical Observatory (SAAO), the Siding Spring Observatory (SSO), and the Cerro Tololo International Observatory (CTIO). For \starA\ and \starB\ we also used data obtained with the Antarctic Search for Transiting ExoPlanets \citep[ASTEP,][]{astep} telescope installed in the Concordia station in the Antartic continent.
Likewise, for \starB\ and \starC\ we also used data obtained with two telescopes (CDK14 0.36 m and the OM-ES) installed in the El Sauce Observatory located in the Rio Hurtado province, in Chile.
Finally, a full transit was obtained for \starC\ with the telescope of the Hazelwood observatory, in Australia.

The photometric observations were reduced and aperture photometry extraction was conducted using \texttt{AstroImageJ} \citep{collins:2017} for all follow-up transit observations except for those of the OM-ES telescope that were based on automated routines previously used in other robotic telescopes \citep[e.g.][]{jordan:2018,brahm:2020}

\begin{figure}[tp]
    \centering
    \includegraphics[width=\columnwidth]{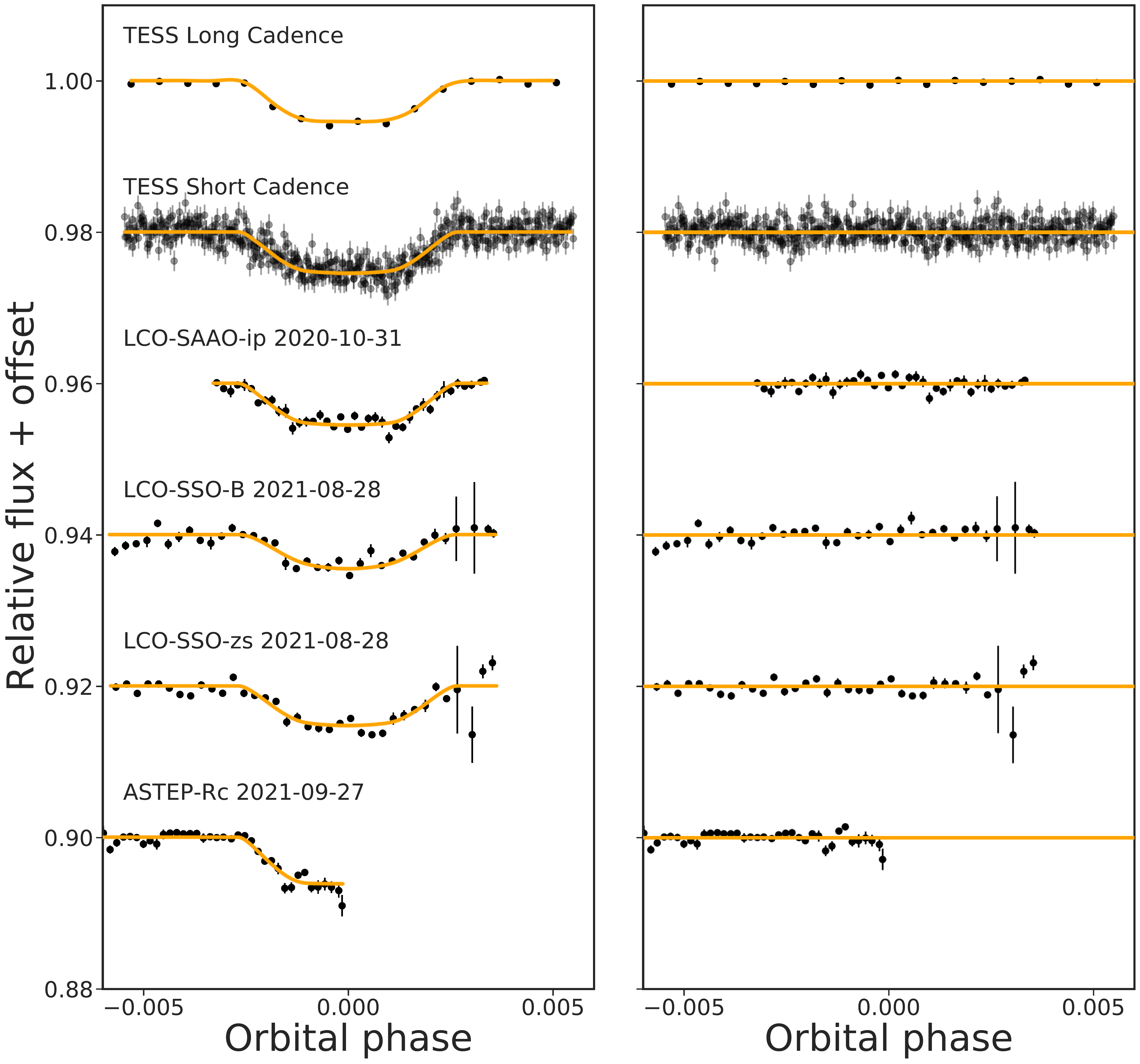}
    \caption{All photometric transits of \plA\ as a function of orbital phase obtained with \tess\ and ground-based facilities. Residuals are presented in the right panel}
    \label{TIC206541859-phot-phase} 
\end{figure}

\begin{figure}[tp]
    \centering
    \includegraphics[width=\columnwidth]{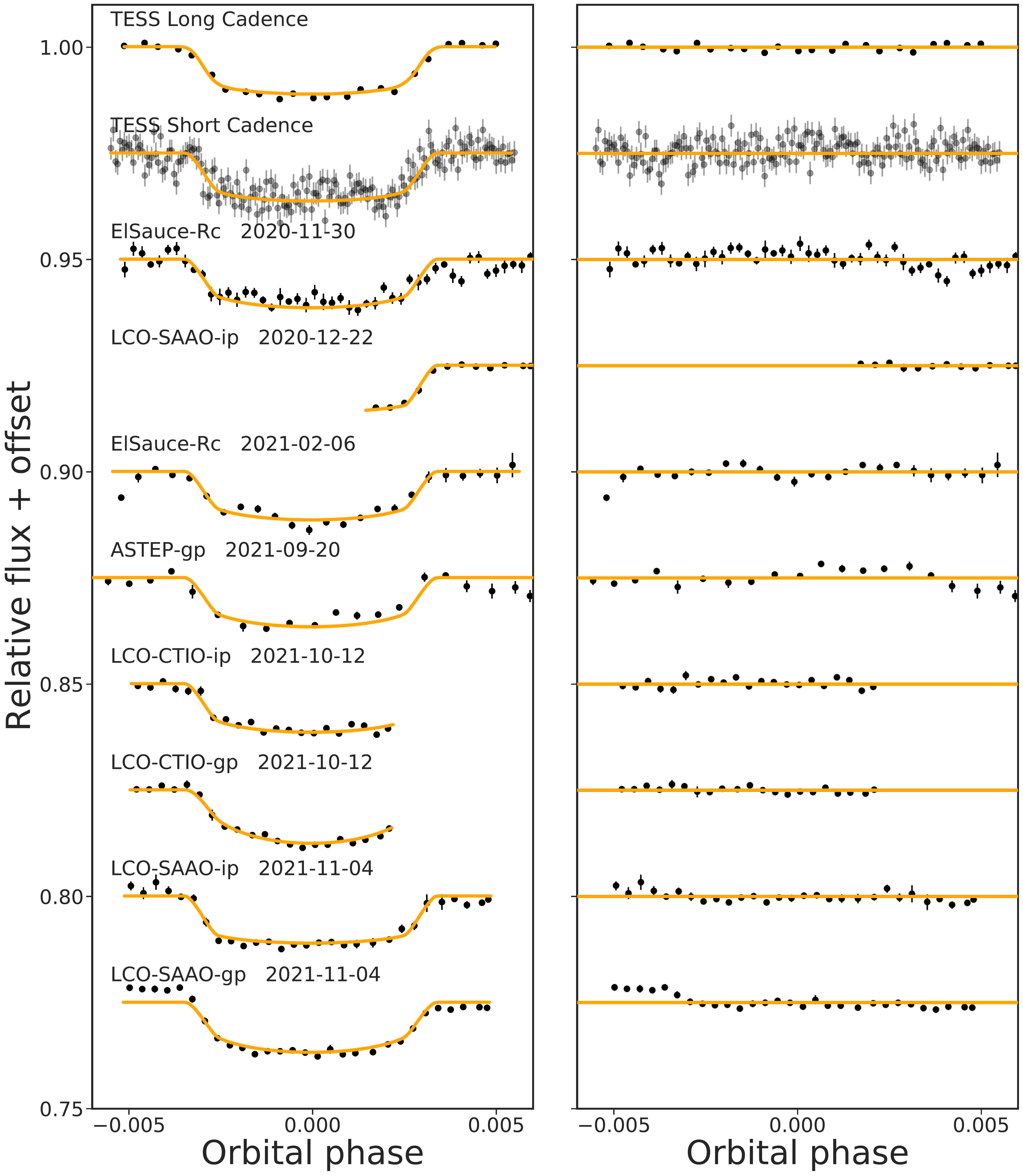}
    \caption{All photometric transits of \plB\ as a function of orbital phase obtained with \tess\ and ground-based facilities. No detrending was performed on the ground-based lightcurves, which generates a negative slope in the residuals of LCO-SAAO.}
    \label{TIC24358417-phot-phase} 
\end{figure}

\begin{figure}[tp]
    \centering
    \includegraphics[width=\columnwidth]{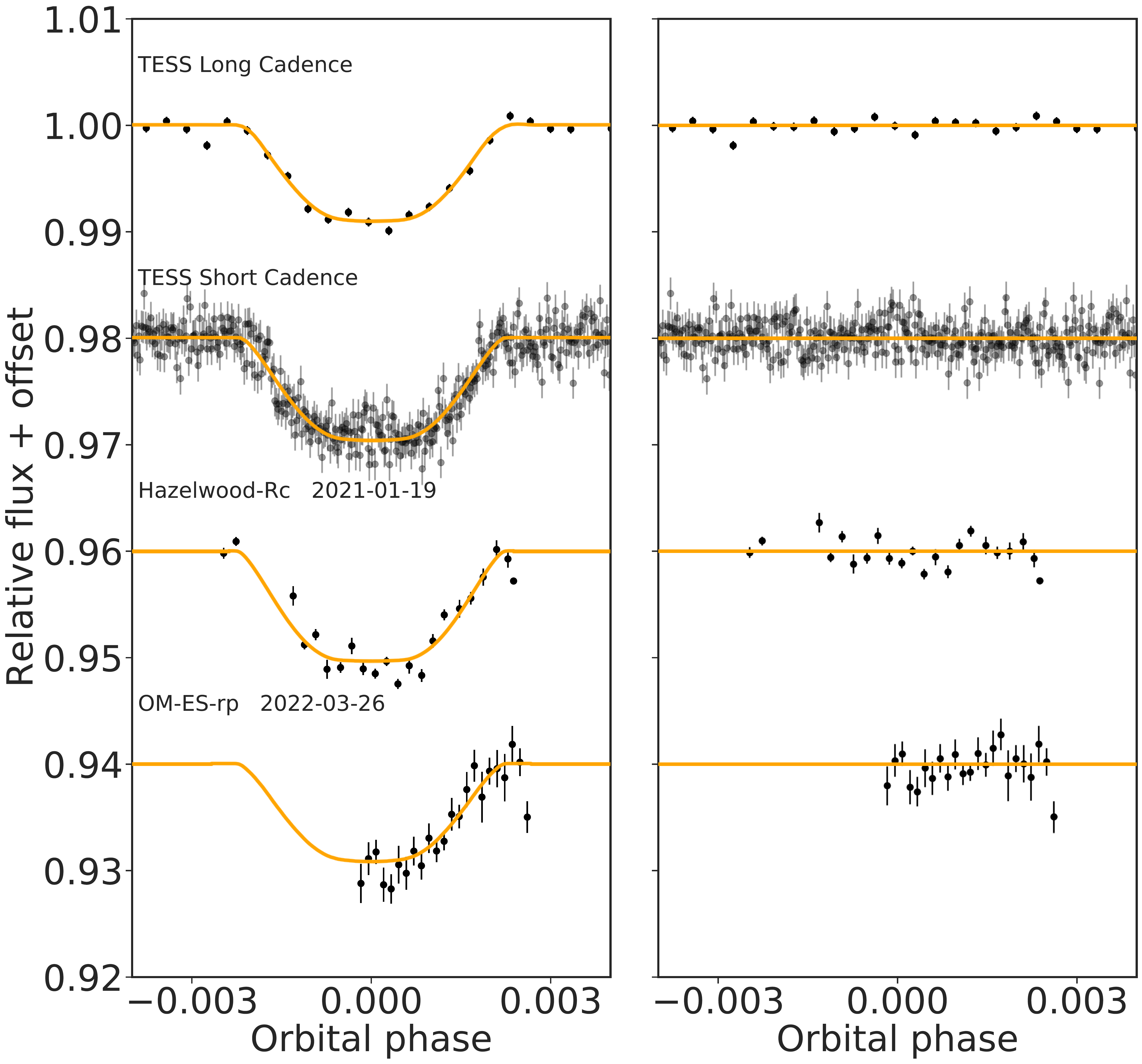}
    \caption{All photometric transits of \plC\ as a function of orbital phase obtained with \tess\ and ground-based facilities.}
    \label{TIC157698565-phot-phase} 
\end{figure}

All ground-based follow-up observations are summarized in Table \ref{tab:gbfu}, where we list the dates of observations, telescope apertures, passband filters, and exposure times.
The light curves\footnote{The data can be downloaded from the ExoFOP platform (\url{https://exofop.ipac.caltech.edu/tess/)}} are also presented in Figures \ref{TIC206541859-phot-phase}, \ref{TIC24358417-phot-phase}, and \ref{TIC157698565-phot-phase}.

\begin{deluxetable*}{lllccc}[]
\tablecaption{Summary of ground based follow-up observations of \starA, \starB, and \starC.  \label{tab:gbfu}}
\tablecolumns{6}
\tablewidth{50pt}
\tablehead{
\colhead{Target} &
\colhead{Observatory} &
\colhead{Date} &
\colhead{Aperture (m)} &
\colhead{Filter} &
\colhead{Exposure time (s)}
\\
}
\startdata
\plA & LCO-SAAO & 2020-10-31 & 1.0 & ip & 60 \\
\plA & LCO-SSO & 2021-08-28 & 1.0  & B  & 25 \\
\plA & LCO-SSO & 2021-08-28 & 1.0  & zs & 37 \\
\plA & ASTEP    & 2021-09-27 & 0.4 & Rc & 60 \\
\hline
\plB & El Sauce & 2020-11-30 & 0.36 & Rc & 90 \\
\plB & LCO-SAAO & 2020-12-22 & 1.0 & ip & 145 \\
\plB & El Sauce & 2021-02-06 & 0.36 & Rc & 180 \\
\plB & ASTEP & 2021-09-20 & 0.4 & gp & 200 \\
\plB & LCO-CTIO & 2021-10-12 & 1.0 & ip & 27 \\
\plB & LCO-CTIO & 2021-10-12 & 1.0 & gp & 41 \\
\plB & LCO-SAAO & 2021-11-04 & 1.0 & gp & 41 \\
\plB & LCO-SAAO & 2021-11-04 & 1.0 & ip & 27 \\
 \hline
\plC & Hazelwood & 2021-01-19 & 0.32 & Rc & 90 \\
\plC & El Sauce (OM-ES) & 2022-03-26 & 0.6 & rp & 9 \\
\enddata
\end{deluxetable*}



\subsection{High Resolution Imaging}
\begin{figure}[tp]
    \centering
    \includegraphics[width=\columnwidth]{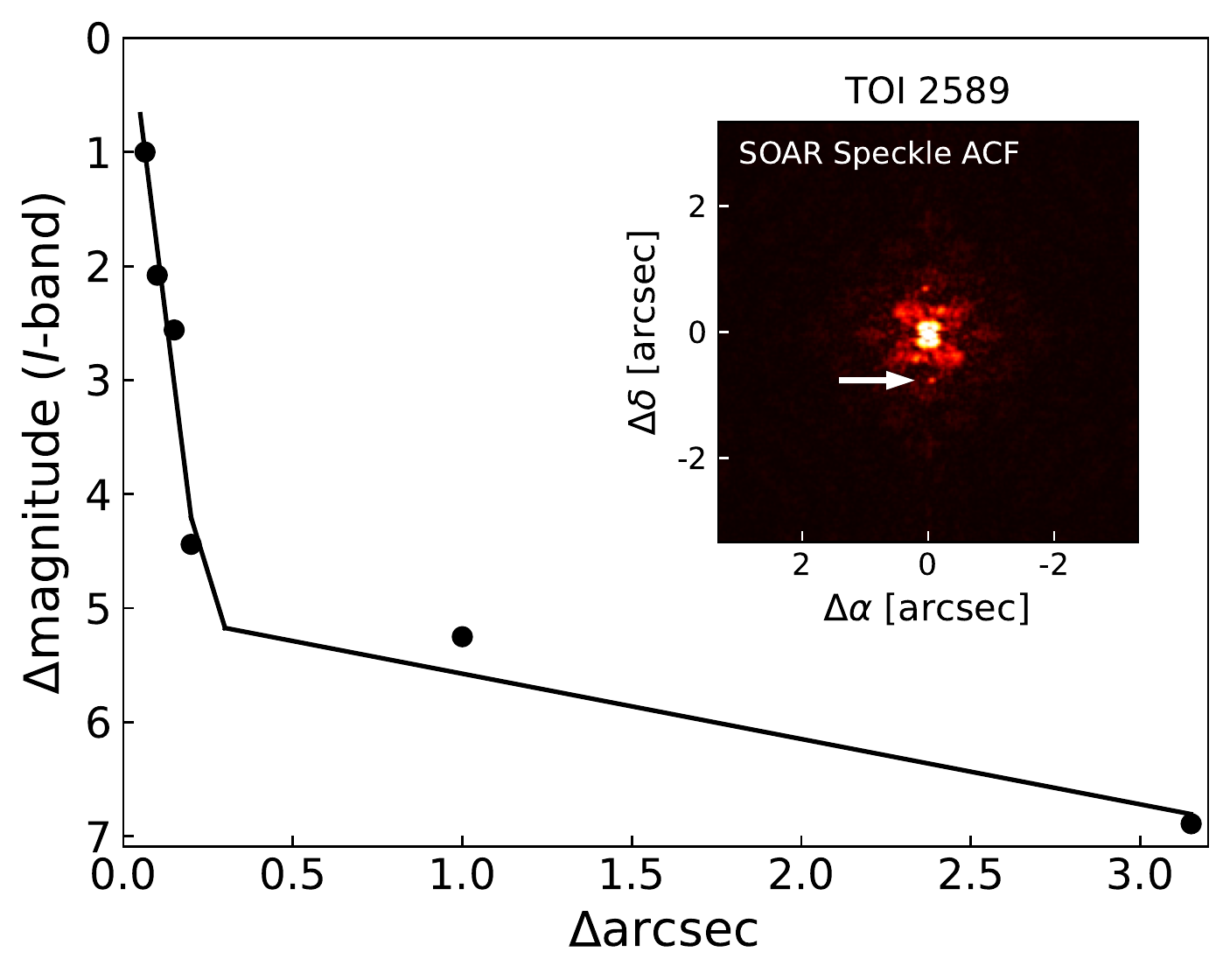}
    \caption{Contrast curve of \starC\ obtained with the Speckle imaging facility HRCam at SOAR. The image inset shows the faint companion star (marked with a white arrow). The additional feature above the star is the same companion mirrored in the auto-correlation function, a product of the speckle image reduction.
}    \label{fig:speckle} 
\end{figure}

We used the optical speckle imaging technique for the three stars presented in this study. Specifically, the three stars were observed with HR Cam of the Southern Astrophysical Research (SOAR) telescope \citep{tokovinin:2018, ziegler:2020}. Observations were performed on 2020-12-03, 2021-10-01, and 2021-11-20 for \starB, \starA, and \starC, respectively. No nearby companions were identified for \starA\ and \starB\ with a contrast of $\Delta I = 7$ mag at 1\arcsec. For \starC\ a faint neighbour ($\Delta I= 3$ mag) at 0.7\arcsec\ was identified (see Figure \ref{fig:speckle}). The flux associated to the neighboring star might impact the determination of the physical parameters of the host star and the planetary transit parameters by diluting the transits. These effects are taken into account in the analysis of the \starB\ system presented in Section \ref{sec:anal}.

\section{Analysis} \label{sec:anal}

\subsection{Stellar Parameters}
\label{sec:ste}

We determined the parameters of the three stars presented in this study following the same iterative procedure adopted in previous discoveries of the WINE collaboration \citep[e.g.][]{schlecker:2020,hobson:2021}. The atmospheric parameters are computed from the co-added FEROS spectra using the \zaspe\ code \citep{brahm:2016:zaspe}, which compares the observed spectra with a grid of synthetic ones. The results of this code are the effective temperature (\teff), surface gravity (\logg), metallicity (\feh), and projected rotational velocity (\vsini) of the corresponding star. These results include uncertainties that include the systematic mismatch between the data and the best fit synthetic model. 
The physical stellar parameters are then computed by comparing public broad-band photometric magnitudes of each star with those produced by the PARSEC stellar evolutionary models \citep{bressan:2012}. In this process we make use of the GAIA DR2 \citep{gaia:dr2} parallax of each star to transform the observed magnitudes to absolute magnitudes. We also fix the metallicity of the stellar models to the value found with \zaspe, while the \zaspe\ effective temperature is used as a prior. We use the \texttt{emcee} package \citep{emcee:2013} to obtain the distributions for the stellar mass (\mstar), stellar age, and interstellar extinction ($A_V$). From these distributions, and using the stellar models we also derive distributions for the stellar radius (\rstar), stellar luminosity (\lstar), stellar density ($\rho_{\star}$), and a new value for the surface gravity (\logg). This new \logg\ value is more constrained than the spectroscopic one, and therefore the iterative process starts by repeating the spectroscopic algorithm fixing the \logg\ value to the one obtained with the evolutionary models, which is followed by a new run of the evolutionary models. We repeat this procedure until convergence is reached for the  \logg\ value, which usually occurs in less than four iterations. The reported stellar uncertainties on the stellar parameters are internal and do not consider possible systematic differences among different stellar evolutionary models, resulting in underestimated uncertainties \citep{tayar:2022}.

We performed a special analysis for \starC\ to incorporate the contamination of the faint companion in the broad-band photometry. First, we considered the faint companion as a gravitational bound object to \starC\ and incorporated a new source with the same metallicity and distance to the analysis presented in the previous paragraph. This analysis did not deliver a good fit to the spectral energy distribution, where the best fit model for the companion star presented significant excess flux in the reddest magnitudes. For this reason, we decided to perform a more detailed analysis where we assumed that the companion star was a chance alignment. To do this we carried out a blend analysis following the procedures described in \citet{2019AJ....157...55H}. We performed a joint modelling of the light curves, RV observations, broad-band photometry measurements, the measured $I$-band magnitude difference between the bright planet host and the neighboring star, the spectroscopically determined stellar atmospheric parameters, and the astrometric parallax from {\em Gaia}. We varied as free parameters the mass, age, metallicity and distance of the planet hosting star, the mass, impact parameter, eccentricity and argument of periastron, and the radius of the planet, and the mass, age, metallicity and distance to the companion star. We also varied the limb darkening coefficients and various light curve detrending parameters. We placed a prior of [Fe/H]$=0.0 \pm 0.5$\,dex on the metallicity of the neighboring star, and set priors on the limb darkening coefficients based on the tabulations from \citet{claret:2012,claret:2013} and \citet{claret:2018}. We used the MIST v1.2 stellar evolution models  \citep{dotter:2016,choi:2016,paxton:2011,paxton:2013,paxton:2015} to determine the radius, luminosity, temperature, and absolute magnitudes of the stars given the varied parameters, and we assumed a line-of-sight Galactic extinction based on the {\sc MWDust} package \citep{bovy:2016}. Based on this analysis, we find that the physical properties of the neighboring star are not very well constrained, but the dilution contributed to each band-pass from the neighbor is fairly well constrained. We find that the neighbor has a mass of $1.34^{+0.43}_{-0.48}$\,M$_{\odot}$, is at a distance modulus of $10.29^{+0.63}_{-1.40}$\,mag, has an age of $0.80^{+2.57}_{-0.61}$\,Gyr, and has a metallicity of [Fe/H]$=-0.86^{+0.45}_{-0.39}$\,dex, where there are significant correlations between these parameters. Having the stellar parameters for the companion star, we included them as an additional source in the PARSEC stellar analysis of \starC\.\

The stellar parameters obtained with the procedure mentioned in the previous paragraphs are presented in Table \ref{tab:stprops}. \starA\ is a late F-type star (\teff = 6219$\pm$70 K), slightly metal rich (\feh = 0.10$\pm$0.05 [dex]), with a mass of \mstar\ = 1.19$\pm$0.03 \msun, and a radius of \rstar\ = 1.29$\pm$0.01 \rsun. On the other hand \starB\ and \starC\ are G-type stars, with effective temperatures of \teff\ = 5581$\pm$60 K and \teff\ = 5552$\pm$70 K, respectively. \starB\ has a metallicity of \feh = 0.22$\pm$0.04 [dex], a mass of \mstar\ = 0.99$\pm$0.03 \msun, and a radius of \rstar\ = 1.05$\pm$0.01 \rsun. \starC\ has a metallicity of \feh = 0.12$\pm$0.04 [dex], a mass of \mstar\ = 0.93$\pm$0.02 \msun, and a radius of \rstar\ = 1.07$\pm$0.01 \rsun. The inclusion of the companion for \starC\ in the analysis changes the stellar mass and radius of the target star by 1 standard deviation.

\begin{deluxetable*}{lrrrc}[b!]
\tablecaption{Stellar properties\tablenotemark{a} of \starA, \starB, and \starC.  \label{tab:stprops}}
\tablecolumns{5}
\tablewidth{0pt}
\tablehead{
\colhead{Parameter} &
\colhead{\starA} &
\colhead{\starB} &
\colhead{\starC\tablenotemark{b}} &
\colhead{Reference} \\
}
\startdata
Names \dotfill   
 & TIC 206541859 & TIC 24358417 & TIC 157698565 & TICv8\\
 & 2MASS J01121157-5655316 & 2MASS J05252264-3440059 & 2MASS J07095718-3713515 & 2MASS  \\
 & TYC 8477-00008-1 & TYC 7063-00698-1 & TYC 7102-00426-1 & TYCHO  \\
RA \dotfill (J2015.5) &  01h12m11.64s & 05h25m22.7s & 07h09m57.16s & Gaia DR2\\
DEC \dotfill (J2015.5) & -56d55m31.39a & -34d40m05.7s & -37d13m51.41s & Gaia DR2\\
pm$^{\rm RA}$ \hfill (mas yr$^{-1}$) & 34.48 $\pm$ 0.04 & 38.00 $\pm$ 0.03 & -18.98 $\pm$ 0.04 & Gaia DR2\\
pm$^{\rm DEC}$ \dotfill (mas yr$^{-1}$) & 19.11 $\pm$ 0.04 & 18.02 $\pm$ 0.03 & 8.32 $\pm$ 0.04 & Gaia DR2\\
$\pi$ \dotfill (mas)& 3.79 $\pm$ 0.03 & 3.16 $\pm$ 0.02 & 4.96 $\pm$ 0.02 & Gaia DR2 \\ 
\hline
T \dotfill (mag) & 10.566 $\pm$ 0.006 & 11.702 $\pm$ 0.006 & 10.721 $\pm$ 0.006 & TICv8\\
B  \dotfill (mag) & 11.61 $\pm$ 0.11 & 13.144 $\pm$ 0.43 & 11.904 $\pm$ 0.18 & APASS\tablenotemark{c}\\
V  \dotfill (mag) & 10.937 $\pm$ 0.007 & 12.483 $\pm$ 0.03 &  11.415 $\pm$ 0.012 & APASS\\
G  \dotfill (mag) & 10.955 $\pm$ 0.008 & 12.186 $\pm$ 0.023 & 11.196  $\pm$ 0.006 & Gaia DR2\tablenotemark{d}\\
G$_{BP}$  \dotfill (mag) & 11.256 $\pm$ 0.005 & 12.581 $\pm$ 0.018 & 11.580 $\pm$ 0.008 & Gaia DR2\\
G$_{RP}$  \dotfill (mag) & 10.519 $\pm$ 0.009 & 11.653 $\pm$ 0.023 & 10.672 $\pm$ 0.009 & Gaia DR2\\
J  \dotfill (mag) &  10.034 $\pm$ 0.024 & 11.076 $\pm$ 0.026 & 10.071 $\pm$ 0.022 & 2MASS\tablenotemark{e}\\
H  \dotfill (mag) &  9.769 $\pm$ 0.025 & 10.746 $\pm$ 0.025 & 9.693 $\pm$ 0.026 & 2MASS\\
K$_s$  \dotfill (mag) & 9.730 $\pm$ 0.023 & 10.664  $\pm$ 0.025 & 9.632 $\pm$ 0.021 & 2MASS\\
\hline
\teff  \dotfill (K) & 6219 $\pm$ 70 & 5581 $\pm$ 60 & 5579 $\pm$ 70 & This work\\
\logg \dotfill (dex) & 4.29 $\pm$ 0.02 & 4.40 $\pm$ 0.02 & $4.35_{-0.01}^{+0.02}$ & This work\\
\feh \dotfill (dex) & 0.10 $\pm$ 0.05 & 0.22 $\pm$ 0.04 & 0.12 $\pm$ 0.04 & This work\\
\vsini \dotfill (km s$^{-1}$) & 4.5 $\pm$ 0.3 & 2.5 $\pm$ 0.3 & 2.2 $\pm$ 0.3 & This work\\
\mstar \dotfill (\msun) & 1.19 $\pm$ 0.03 & $0.99_{-0.02}^{+0.03}$ & $0.93_{-0.02}^{+0.03}$ & This work\\
\rstar \dotfill (\rsun) & 1.29 $\pm$ 0.01 & 1.05 $\pm$ 0.01 & 1.07 $\pm$ 0.01 & This work\\
L$_{\star}$ \dotfill (L$_{\odot}$) & 2.3 $\pm$ 0.1 & $0.97_{-0.03}^{+0.04}$ &  $0.99_{-0.03}^{+0.04}$ & This work\\
en qA$_{V}$ \dotfill (mag) & 0.17 $\pm$ 0.06 & $0.07_{-0.05}^{+0.06}$ &  $0.11_{-0.05}^{+0.06}$ & This work\\
Age \dotfill (Gyr) & 2.9 $\pm$ 0.7 & 7 $\pm$ 2 & $11_{-2}^{+2}$ &This work\\
$\rho_\star$ \dotfill (g cm$^{-3}$) & 0.78 $\pm$ 0.03 & 1.22 $\pm$ 0.07  &  1.07 $\pm$ 0.07  & This work\\
\enddata
\tablecomments{$^a$ The stellar parameters computed in this work do not consider possible systematic differences among different stellar evolutionary models \citep{tayar:2022} and have underestimated uncertainties. $^b$We note that the magnitudes reported for \starC\ include the contamination of a close neighbor ($\Delta I = 2.9$ mag),$^c$\citet{apass},$^d$\citet{gaia:dr2},$^e$\citet{2mass}}.
\end{deluxetable*}

\subsection{Global Modelling}

For each system we simultaneously model \tess\ photometry, ground based light curves, and precision radial velocities with the \texttt{juliet} package \citep{espinoza:2019}. \texttt{juliet} searches the global posterior maximum by computing the Bayesian log evidence ($\ln{\zeta}$) with a nested-sampling algorithm.
\texttt{juliet} uses the \texttt{radvel} package \citep{fulton:2018} to model the radial velocities, while the photometric transits are modelled with \texttt{batman} \citep{kreidberg:2015}. In this study we use the \texttt{dynesty} \citep{dynesty} package incorporated in \texttt{juliet} to perform the posterior sampling with 500 live points.  
For \tess\ 30-minute cadence data we used the oversampling technique to model the transit light curves \citep{kipping:2010}.
For the three systems presented in this analysis we adopt the stellar density as a free parameter of the transit model instead of the $a/R_{\star}$ parameter, and we use the density derived from the stellar analysis as a gaussian prior for this parameter. We note that the stellar parameters do not include uncertainty from systematic differences among different stellar evolutionary models, resulting in underestimated uncertainties on planetary parameters.

\subsubsection{Global Modelling of \plA}
For the analysis of \plA\, we use the 30-minute cadence light curve of sector 2 computed with \texttt{tesseract}, the PDCSAP light curves of sectors 28 and 29, and all four ground based photometric light curves. For \tess\ we adopt the quadratic limb darkening law, where both parameters were set as free parameters with uniform priors between 0 and 1. For the ground based light curves, we adopt the linear limb darkening law with uniform priors between 0 and 1 for each telescope. We also include a Gaussian process to model systematic variations present in the 30-minute cadence light curve of sector 2, for which we use a Matern 3/2 Kernel.
In addition to the photometry, we use the CHIRON, CORALIE, FEROS, and HARPS radial velocities as data, and model them with a simple Keplerian model. We also consider distinct radial velocity zero points and jitter factors for each instrument as free parameters. We tried two different models for \plA\ with \texttt{juliet}. First we forced a circular orbit by fixing the eccentricity to 0, and then we considered a model with the eccentricity and argument of the periastron as free parameters. Even though the eccentricity is not strongly constrained, the eccentric model was strongly favoured (seventeen times more likely) when comparing the Bayesian log-evidences of both models.

The prior distributions along with the resulting parameters for the adopted model for \plA\ are presented in Table \ref{tab:plpropsA}. \plA\ is a warm giant planet with a mass of \mpl$=0.30_{-0.03}^{+0.02}$ \mjup\ and a radius of \rpl$=1.00_{-0.02}^{+0.02}$ \rjup. It has a low eccentricity orbit ($e=0.15\pm0.05$) with a period of 30.0836 days, and a time averaged equilibrium temperature of $903\pm 15$ K \citep[assuming zero albedo and uniform planet surface temperature,][]{mendez:2017}.

\subsubsection{Global Modelling of \plB}
For the global modelling of \plB\, we used the \tess\ 30-minute cadence light curves of sectors 5 and 6 and the 2-minute cadence light curve of sector 32, as well as the ground-based follow-up light curves. To model the \tess\ light curves we adopted the quadratic limb darkening law, while ground-based light curves were modeled with the linear limb darkening law. Two distinct Gaussian processes were adopted for modeling the \tess\ light curves of sectors 5 and 6. A Matern 3/2 kernel was used in both cases. We modeled the HARPS and FEROS radial velocities of \starB\ with an eccentric Keplerian signal. Distinct zero points and jitter terms were used for modeling each instrument.

The prior distributions along with the resulting parameters for the adopted model of \plB\ are presented in Table \ref{tab:plpropsB}. \plB\ is a massive warm giant planet with a mass of \mpl$=5.98_{-0.21}^{+0.020}$ \mjup\ and a radius of \rpl$=1.00_{-0.02}^{+0.02}$ \rjup. It has highly eccentric orbit of $e= 0.676 \pm 0.002$ with a period of $22.65398\pm0.00002$ days, and a time averaged equilibrium temperature of $799\pm 10$ K.

\subsubsection{Global Modelling of \plC}

For the global modelling of \plC\, we adopted the quadratic limb darkening law for the \tess\ light curves of sectors 7 and 34, and the linear limb darkening law for the ground based light curves. \tess\ data was also modelled with a distinct Gaussian process for each sector with a Matern 3/2 Kernel. For the ground based light curve obtained with the Hazelwood Observatory we included a linear model where the airmass was user as the linear regressor. For modelling the radial velocities we adopted an eccentric Keplerian signal, with different radial velocity zero points and jitter terms for each instrument. For the ground-based light curves and the \tess\ light curve of sector 34 we included the contamination of the faint companion by considering fixed dilution factors of 0.92 and 0.93, respectively. These dilution factors were computed from the magnitudes of the faint companion star obtained from the analysis presented in section \ref{sec:ste}. For the 
\tess\ light curve of sector 7, we adopted a uniform prior for the dilution factor to allow for additional contaminating sources.

The prior distributions along with the resulting parameters for the adopted model of \plC\ are presented in Table \ref{tab:plpropsC}. \plC\ is a long period ($P=61.6277\pm0.0002$ d) warm Jupiter with a mass of \mpl$=3.5_{-0.1}^{+0.01}$ \mjup\ and a radius of \rpl$=1.09_{-0.03}^{+0.03}$ \rjup. It has a moderately large orbital eccentricity of $e=0.523\pm0.006$ and a time averaged equilibrium temperature of $593\pm 8$ K.

\begin{deluxetable}{lrr}
\tablecaption{Prior and posterior parameters of the global analysis of \plA. For the priors, $N(\mu,\sigma)$ stands for a normal distribution with mean $\mu$ and standard deviation $\sigma$, $U(a,b)$ stands for a uniform distribution between $a$ and $b$, and $LU(a,b)$ stands for a log-uniform prior defined between $a$ and $b$. The stellar parameters, from which the planetary parameters are derived, do not consider possible systematic differences among different stellar evolutionary models \citep{tayar:2022} and have underestimated uncertainties.
\label{tab:plpropsA}
}
\tabletypesize{\footnotesize}
\tablecolumns{3}
\tablewidth{0pt}
\tablehead{
\colhead{Parameter} &
\colhead{Prior} &
\colhead{Value}
}
\startdata
P (days) & $N(30.1,0.1)$  &          $30.08364_{-0.00005}^{+0.00005}$  \\
T$_0$ (BJD)&  $N(2458372.19,0.01)$&  $2458372.194_{0.001}^{0.001}$ \\
$\rho_{\star}$ (Kg m$^{-3}$) & $N(776,60)$ & $784_{-49}^{+48}$ \\
\rpl/\rstar  & $U(0.001,1)$ & $0.079_{-0.001}^{+0.001}$ \\
b & $U(0,1)$ & $0.887_{-0.008}^{+0.008}$ \\
K (km s$^{-1}$) & $U(0,0.2)$& $0.017_{-0.002}^{+0.002}$ \\
$e$ & $U(0,0.9)$ & $0.15_{-0.04}^{+0.05}$ \\
$\omega$ & $(0,360)$ & $39_{-15}^{+18}$ \\ 
q$_1^{\rm TESS}$ & $U(0,1)$ & $0.3_{-0.1}^{+0.1}$ \\
q$_2^{\rm TESS}$ & $U(0,1)$ & $0.6_{-0.3}^{+0.2}$ \\
q$_1^{\rm LCO-SAAO-ip}$  & $U(0,1)$ & $0.2_{-0.1}^{+0.1}$ \\
q$_2^{\rm LCO-SAAO-ip}$  & $U(0,1)$ & $0.7_{-0.3}^{+0.2}$ \\
q$_1^{\rm LCO-SSO-zs}$  & $U(0,1)$ & $0.6_{-0.1}^{+0.1}$ \\
q$_1^{\rm LCO-SSO-B}$  & $U(0,1)$ & $0.94_{-0.07}^{+0.04}$ \\
q$_1^{\rm ASTEP-Rc}$  & $U(0,1)$ & $0.1_{-0.1}^{+0.1}$ \\
$\sigma_w^{\rm TESS,S2}$ (ppm)  & $LU(10^{-1},10^3)$ & $204_{-15}^{+14}$ \\
$\sigma_w^{\rm TESS,S28}$ (ppm) & $LU(10^{-1},10^3)$ & $3_{-2}^{+17}$ \\
$\sigma_w^{\rm TESS,S29}$ (ppm) & $LU(10^{-1},10^3)$ & $8_{-7}^{+63}$\\
$\sigma_w^{\rm LCO-SAAO-ip}$ (ppm) & $LU(10^{-1},10^4)$ & $1190_{-110}^{+110}$ \\
$\sigma_w^{\rm LCO-SSO-zs}$ (ppm) & $LU(10^{-1},10^4)$ & $1250_{-130}^{+140}$ \\
$\sigma_w^{\rm LCO-SSO-B}$ (ppm) & $LU(10^{-1},10^4)$ & $1410_{-120}^{+140}$ \\
$\sigma_w^{\rm ASTEP-Rc}$ (ppm) & $LU(10^{-1},10^4)$ & $610_{-520}^{+210}$ \\
$\mu_{\rm flux}^{\rm TESS,28}$ (ppm) & $N(0,0.1)$ &   $-0.00002_{-0.00001}^{+0.00001}$ \\
$\mu_{\rm flux}^{\rm TESS,29}$ (ppm) & $N(0,0.1)$ & $-0.00003_{-0.00001}^{+0.00001}$ \\
$\mu_{\rm flux}^{\rm LCO-SAAO-ip}$ (ppm) & $N(0,0.1)$ & $-0.002_{-0.0002}^{+0.0002}$ \\
$\mu_{\rm flux}^{\rm LCO-SSSO-zs}$ (ppm) & $N(0,0.1)$ & $-0.0017_{-0.0002}^{+0.0002}$ \\
$\mu_{\rm flux}^{\rm LCO-SSO-B}$ (ppm) & $N(0,0.1)$ & $-0.0014_{-0.0002}^{+0.0002}$ \\
$\mu_{\rm flux}^{\rm ASTEP-Rc}$ (ppm) & $N(0,0.1)$ & $-0.0008_{-0.0001}^{+0.0001}$ \\
$\gamma_{\rm CHIRON}$  (km s$^{-1}$)& $N(0.00,0.03)$ & $-0.01_{-0.01}^{+0.01}$ \\
$\gamma_{\rm CORALIE}$  (km s$^{-1}$) & $N(24.19,0.03)$ & $24.184_{-0.007}^{+0.007}$ \\
$\gamma_{\rm FEROS}$  (km s$^{-1}$) & $N(24.18,0.03)$ & $24.191_{-0.004}^{+0.004}$ \\
$\gamma_{\rm HARPS}$ (km s$^{-1}$) & $N(24.18,0.03)$ & $24.187_{-0.002}^{+0.002}$ \\
$\sigma_{\rm CHIRON}$   (km s$^{-1}$)& $N(0.001,0.1)$ & $0.01_{-0.01}^{+0.02}$ \\
$\sigma_{\rm CORALIE}$  (km s$^{-1}$)& $N(0.001,0.1)$ & $0.002_{-0.001}^{+0.005}$ \\
$\sigma_{\rm FEROS}$ (km s$^{-1}$)& $N(0.001,0.1)$ & $0.015_{-0.003}^{+0.004}$ \\
$\sigma_{\rm HARPS}$  (km s$^{-1}$)& $N(0.001,0.1)$& $0.004_{-0.001}^{+0.002}$ \\
$\sigma^{GP}_{\rm TESS,2}$ &  $LU(10^{-5},10^{5})$ & $-0.0008_{-0.0001}^{+0.0001}$ \\
$\rho^{GP}_{\rm TESS,2}$ &  $LU(10^{-3},10^{3})$ & $2.0_{-0.4}^{+0.5}$ \\
\hline
$a/$\rstar    & & 33.5$^{+0.7}_{-0.7}$ \\
$i$ (deg) &  & 88.48$^{+0.04}_{-0.04}$ \\
\mpl\ (\mjup)& &  0.30$_{-0.03}^{+0.03}$ \\
\rpl\ (\rjup)& & 1.00$_{-0.02}^{+0.02}$ \\
$a$ (AU)    & & 0.201$^{+0.005}_{-0.005}$ \\
\teq (K)  & & 904$^{+16}_{-17}$ \\ 
\enddata
\end{deluxetable}

\begin{deluxetable}{lrr}
\tablewidth{\columnwidth}
\tablecaption{Prior and posterior parameters of the global analysis of \plB. See Table \ref{tab:plpropsA} for the definition of prior distributions. The stellar parameters, from which the planetary parameters are derived, do not consider possible systematic differences among different stellar evolutionary models \citep{tayar:2022} and have underestimated uncertainties. \label{tab:plpropsB}}
\tabletypesize{\footnotesize}
\tablecolumns{3}
\tablewidth{0pt}
\tablehead{
\colhead{Parameter} &
\colhead{Prior} &
\colhead{Value}
}
\startdata
P (days)    &  $N(22.7,0.1)$      &  $22.65398_{-0.00002}^{+0.00002}$  \\
T$_0$ (BJD) &  $N(2458458.8,0.2)$ &  $2458458.7545_{-0.0008}^{+0.0008}$ \\
$\rho_{\star}$ (Kg m$^{-3}$)      & $N(1120,90)$ & $1301_{-58}^{+51}$ \\
\rpl/\rstar  & $U(0.001,1)$       & $0.0980_{-0.0008}^{+0.0008}$ \\
b & $U(0,1)$ & $0.26_{-0.06}^{+0.06}$ \\
K (km s$^{-1}$) & $U(0.1,2.0)$ & $0.584_{-0.004}^{+0.004}$ \\
$e$ & $U(0,0.9)$ & $0.676_{-0.002}^{+0.002}$ \\
$\omega$ (deg) & $(0,360)$ & $12.6_{-0.6}^{+0.7}$ \\ 
q$_1^{\rm TESS}$ & $U(0,1)$ & $0.4_{-0.2}^{+0.2}$ \\
q$_2^{\rm TESS}$ & $U(0,1)$ & $0.2_{-0.1}^{+0.2}$ \\
q$_1^{\rm LCO-SAAO-ip}$  & $U(0,1)$ & $0.4_{-0.1}^{+0.1}$ \\
q$_1^{\rm ElSauce-Rc}$  & $U(0,1)$ & $0.45_{-0.08}^{+0.08}$ \\
q$_1^{\rm ASTEP-Rc}$  & $U(0,1)$ & $0.5_{-0.1}^{+0.1}$ \\
q$_1^{\rm LCO-SAAO-0.4m-ip}$  & $U(0,1)$ & $0.47_{-0.08}^{+0.07}$ \\
q$_1^{\rm LCO-SAAO-0.4m-gp}$  & $U(0,1)$ & $0.88_{-0.05}^{+0.05}$ \\
q$_1^{\rm LCO-CTIO-ip}$  & $U(0,1)$ & $0.5_{-0.1}^{+0.1}$ \\
q$_1^{\rm LCO-CTIO-gp}$  & $U(0,1)$ & $0.73_{-0.06}^{+0.06}$ \\
$\sigma_w^{\rm TESS,S5}$ (ppm)  & $LU(10^{-1},10^3)$ & $370_{-30}^{+30}$ \\
$\sigma_w^{\rm TESS,S6}$ (ppm) & $LU(10^{-1},10^3)$ & $3_{-2}^{+17}$ \\
$\sigma_w^{\rm TESS,S32}$ (ppm) & $LU(10^{-1},10^3)$ & $4_{-4}^{+32}$\\
$\sigma_w^{\rm LCO-SAAO-ip}$ (ppm)      & $LU(10^{-1},10^4)$ & $590_{-90}^{+100}$ \\
$\sigma_w^{\rm ElSauce-Rc-1}$ (ppm) & $LU(10^{-1},10^4)$ & $10_{-10}^{+150}$ \\
$\sigma_w^{\rm ElSauce-Rc-2}$ (ppm) & $LU(10^{-1},10^4)$ & $70_{-70}^{+940}$ \\
$\sigma_w^{\rm ASTEP-Rc}$ (ppm) & $LU(10^{-1},10^4)$ & $2260_{-180}^{+200}$ \\
$\sigma_w^{\rm LCO-SAAO-0.4m-ip}$ (ppm) & $LU(10^{-1},10^4)$ & $1120_{-130}^{+150}$ \\
$\sigma_w^{\rm LCO-SAAO-0.4m-gp}$ (ppm) & $LU(10^{-1},10^4)$ & $1100_{-110}^{+120}$ \\
$\sigma_w^{\rm LCO-CTIO-ip}$ (ppm) & $LU(10^{-1},10^4)$ & $1760_{-190}^{+200}$ \\
$\sigma_w^{\rm LCO-CTIO-gp}$ (ppm) & $LU(10^{-1},10^4)$ & $1490_{-140}^{+150}$ \\
$\mu_{\rm dil}^{\rm TESS,S5}$ & $U(0.5,1.0)$ &   $0.986_{-0.016}^{+0.009}$ \\
$\mu_{\rm dil}^{\rm TESS,S6}$ & $U(0.5,1.0)$ &   $0.989_{-0.012}^{+0.008}$ \\
$\mu_{\rm flux}^{\rm TESS,S32}$ (ppm) & $N(0,0.1)$ &   $0.00001_{-0.00002}^{+0.00002}$ \\
$\mu_{\rm flux}^{\rm LCO-SAAO-ip}$ (ppm) & $N(0,0.1)$ & $-0.0008_{-0.0001}^{+0.0001}$ \\
$\mu_{\rm flux}^{\rm ElSauce-Rc-1}$ (ppm) & $N(0,0.1)$ & $-0.004_{-0.0003}^{+0.0003}$ \\
$\mu_{\rm flux}^{\rm ElSauce-Rc-2}$ (ppm) & $N(0,0.1)$ & $-0.0061_{-0.0002}^{+0.0002}$ \\
$\mu_{\rm flux}^{\rm ASTEP-Rc}$ (ppm) & $N(0,0.1)$ & $-0.0045_{-0.0002}^{+0.0002}$ \\
$\mu_{\rm flux}^{\rm LCO-SAAO-0.4m-ip}$ (ppm) & $N(0,0.1)$ & $-0.0095_{-0.0002}^{+0.0002}$ \\
$\mu_{\rm flux}^{\rm LCO-SAAO-0.4m-gp}$ (ppm) & $N(0,0.1)$ & $-0.0105_{-0.0002}^{+0.0002}$ \\
$\mu_{\rm flux}^{\rm LCO-CTIO-ip}$ (ppm) & $N(0,0.1)$ & $-0.0077_{-0.0002}^{+0.0002}$ \\
$\mu_{\rm flux}^{\rm LCO-CTIO-gp}$ (ppm) & $N(0,0.1)$ & $-0.0080_{-0.0002}^{+0.0002}$ \\
$\sigma_{\rm FEROS}$ (km s$^{-1}$) & $N(0.001,0.1)$  & $0.020_{-0.004}^{+0.006}$ \\
$\sigma_{\rm HARPS}$ (km s$^{-1}$) & $N(0.001,0.1)$  & $0.003_{-0.002}^{+0.003}$ \\
$\gamma_{\rm FEROS}$ (km s$^{-1}$) & $N(83.92,0.03)$ & $83.926_{-0.005}^{+0.006}$ \\
$\gamma_{\rm HARPS}$ (km s$^{-1}$) & $N(83.95,0.03)$ & $83.942_{-0.002}^{+0.002}$ \\
\hline
$a/$\rstar     & & 32.3$^{+0.6}_{-0.6}$ \\
$i$ (deg)      & & 89.52$^{+0.02}_{-0.02}$ \\
\mpl\ (\mjup)  & & 5.98$_{-0.20}^{+0.21}$ \\
\rpl\ (\rjup)  & & 1.00$_{-0.02}^{+0.02}$ \\
$a$ (AU)       & & 0.158$^{+0.003}_{-0.003}$ \\
\teq (K)       & & 799$^{+10}_{-11}$ \\ 
\enddata
\end{deluxetable}

\begin{deluxetable}{lrr}
\tablecaption{Prior and posterior parameters of the global analysis of \plC . See Table \ref{tab:plpropsA} for the definition of prior distributions. The stellar parameters, from which the planetary parameters are derived, do not consider possible systematic differences among different stellar evolutionary models \citep{tayar:2022} and have underestimated uncertainties. \label{tab:plpropsC}}
\tabletypesize{\footnotesize}
\tablecolumns{3}
\tablewidth{0pt}
\tablehead{
\colhead{Parameter} &
\colhead{Prior} &
\colhead{Value}
}
\startdata
P (days)    &  $N(61.6,0.5)$      &  $61.6277_{-0.0002}^{+0.0002}$  \\
T$_0$ (BJD) &  $N(2458494.6,0.2)$ &  $2458494.579_{-0.002}^{+0.002}$ \\
$\rho_{\star}$ (Kg m$^{-3}$)      & $N(1070,70)$ & $1140_{-50}^{+50}$ \\
\rpl/\rstar  & $U(0.001,1)$       & $0.107_{-0.003}^{+0.003}$ \\
b & $U(0,1)$ & $0.878_{-0.008}^{+0.006}$ \\
K (km s$^{-1}$) & $U(0.15,0.30)$ & $0.221_{-0.003}^{+0.003}$ \\
$e$ & $U(0,0.9)$ & $0.522_{-0.006}^{+0.006}$ \\
$\omega$ & $(0,360)$ & $218_{-1}^{+1}$ \\ 
q$_1^{\rm TESS}$ & $U(0,1)$ & $0.14_{-0.08}^{+0.10}$ \\
q$_2^{\rm TESS}$ & $U(0,1)$ & $0.8_{-0.2}^{+0.2}$ \\
q$_1^{\rm HWD}$  & $U(0,1)$ & $0.2_{-0.1}^{+0.2}$ \\
q$_1^{\rm OM-ES}$  & $U(0,1)$ & $0.5_{-0.3}^{+0.3}$ \\
$\sigma_w^{\rm TESS,S7}$ (ppm)  & $LU(10^{-1},10^3)$ & $480_{-20}^{+20}$ \\
$\sigma_w^{\rm TESS,S34}$ (ppm) & $LU(10^{-1},10^3)$ & $4_{-4}^{+50}$ \\
$\sigma_w^{\rm HWD}$ (ppm)      & $LU(10^{-1},10^4)$ & $1890_{-140}^{+150}$ \\
$\sigma_w^{\rm OM-ES}$ (ppm)      & $LU(10^{-1},10^4)$ & $10_{-10}^{+220}$ \\
$\mu_{\rm flux}^{\rm HWD}$ (ppm) & $N(0,0.1)$ & $-0.015_{-0.002}^{+0.002}$ \\
$\mu_{\rm flux}^{\rm OM-ES}$ (ppm) & $N(0,0.1)$ & $-0.006_{-0.001}^{+0.001}$ \\
$\mu_{\rm dil}^{\rm TESS,S7}$  & $U(0.7,1.0)$ &   $0.94_{-0.03}^{+0.03}$ \\
$\sigma_{\rm CHIRON}$ (km s$^{-1}$) & $LU(0.001,0.1)$  & $0.018_{-0.005}^{+0.007}$ \\
$\sigma_{\rm FEROS}$ (km s$^{-1}$) & $LU(0.001,0.1)$  & $0.015_{-0.003}^{+0.003}$ \\
$\sigma_{\rm HARPS}$ (km s$^{-1}$) & $LU(0.001,0.1)$  & $0.010_{-0.002}^{+0.002}$ \\
$\sigma_{\rm CORALIE}$ (km s$^{-1}$) & $LU(0.001,0.1)$  & $0.001_{-0.0008}^{+0.0046}$ \\
$\gamma_{\rm CHIRON}$ (km s$^{-1}$) & $N(-1.80,0.05)$ & $-1.805_{-0.007}^{+0.007}$ \\
$\gamma_{\rm FEROS}$ (km s$^{-1}$) & $N(68.55,0.03)$ & $68.549_{-0.004}^{+0.004}$ \\
$\gamma_{\rm HARPS}$ (km s$^{-1}$) & $N(68.58,0.03)$ & $65.579_{-0.002}^{+0.002}$ \\
$\gamma_{\rm CORALIE}$ (km s$^{-1}$) & $N(68.58,0.03)$ & $68.566_{-0.009}^{+0.009}$ \\
$\sigma^{GP}_{\rm TESS,S7}$ & $LU(10^{-5},10^{4})$ & $0.0009_{-0.0003}^{+0.006}$ \\
$\rho^{GP}_{\rm TESS,S7}$   & $LU(10^{-5},10^{4})$ & $4_{-2}^{+3}$ \\
$\sigma^{GP}_{\rm TESS,S34}$ & $LU(10^{-5},10^{4})$ & $0.00019_{-0.00004}^{+0.00006}$ \\
$\rho^{GP}_{\rm TESS,S34}$   & $LU(10^{-5},10^{4})$ & $2.2_{-0.6}^{+0.9}$ \\
$\theta_{0}^{\rm HWD}$   & $U(-1,1)$ & $-0.007_{-0.001}^{+0.001}$ \\
\hline
$a/$\rstar     & & 61.2$^{+0.9}_{-0.9}$ \\
$i$ (deg)      & & 89.17$^{+0.01}_{-0.01}$ \\
\mpl\ (\mjup)  & & 3.5$_{-0.1}^{+0.1}$ \\
\rpl\ (\rjup)  & & 1.08$_{-0.03}^{+0.03}$ \\
$a$ (AU)       & & 0.300$^{+0.006}_{-0.005}$ \\
\teq (K)       & & 592$^{+7}_{-8}$ \\ 
\enddata
\end{deluxetable}

 \begin{figure*}[tp]
    \centering
    \includegraphics[width=\textwidth]{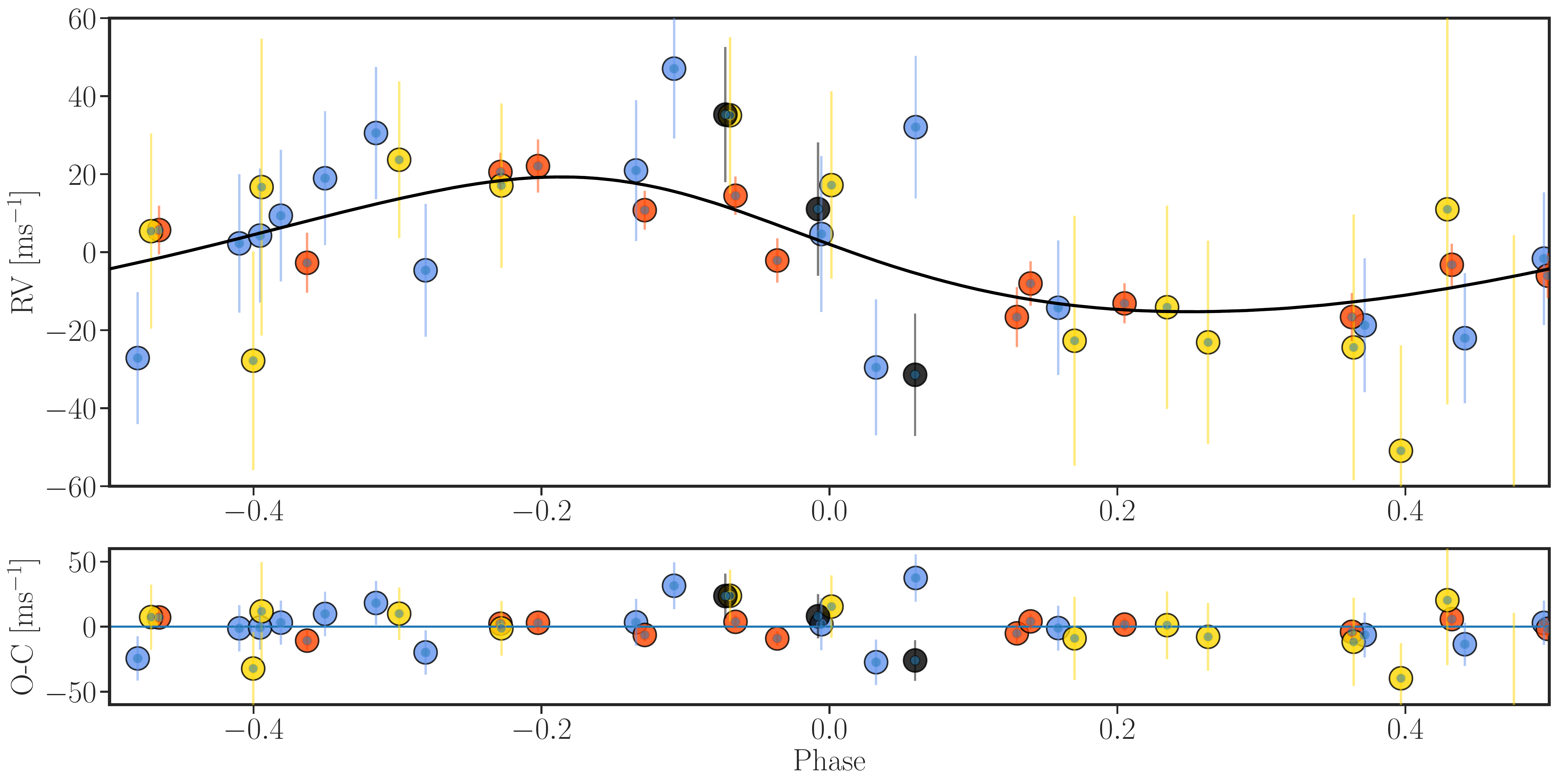}
    \caption{Radial velocities of \starA\ as a function of the orbital phase of \plA\ (FEROS: blue, Coralie: yellow, CHIRON: black, HARPS: orange). The solid line corresponds to the Keplerian model using the posterior parameters of the joint fit.}
    \label{TIC206541859-rvs-phase} 
\end{figure*}

 \begin{figure*}[tp]
    \centering
    \includegraphics[width=\textwidth]{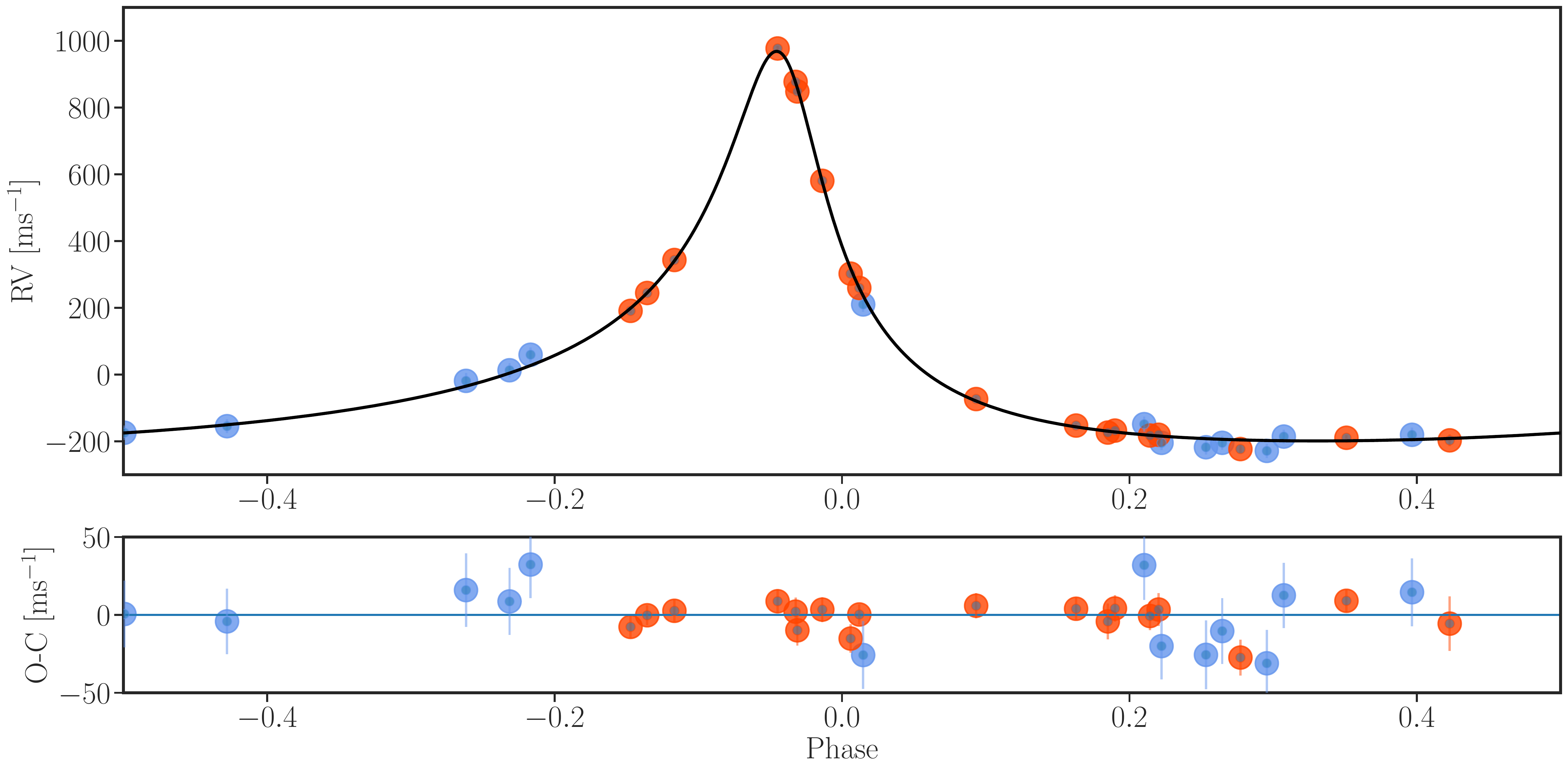}
    \caption{Radial velocities of \starB\ as a function of the orbital phase of \plB\ (FEROS: blue, HARPS: orange). The solid line corresponds to the Keplerian model using the posterior parameters of the joint fit.}
    \label{rvs} 
\end{figure*}

 \begin{figure*}[tp]
    \centering
    \includegraphics[width=\textwidth]{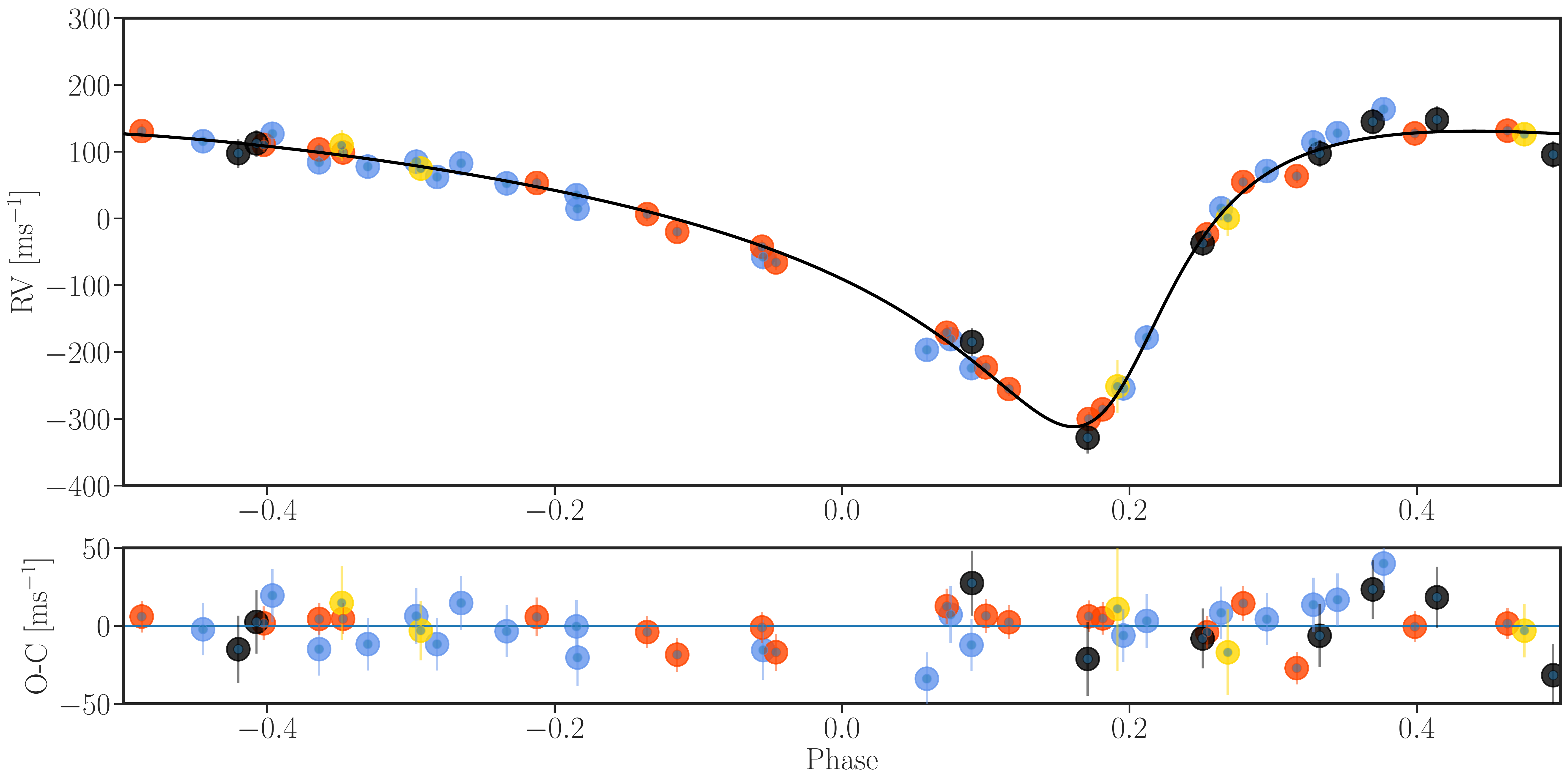}
    \caption{Radial velocities of \starC\ as a function of the orbital phase of \plC\ (FEROS: blue, CHIRON: black, HARPS: orange, CORALIE: yellow). The solid line corresponds to the Keplerian model using the posterior parameters of the joint fit.}
    \label{TIC157698565-rvs-phase} 
\end{figure*}

\subsection{Search for additional signals}

We searched for additional periodic signals in the \tess\ light curves of the three stars presented in this study. For each Presearch Data Conditioning Simple Aperture Photometry \citep[PDCSAP]{smith:2012,stumpe:2014} light curve with a cadence of 120s, we removed outliers, masked the known transits and applied a Generalized Lomb-Scargle Periodogram \citep[GLS,][]{zechmeister:2009} to identify possible sinusoidal variations produced by the rotation of the stars. We found no significant signals for any of the three systems. After that we also applied a Box-Least Squares \citep[BLS,][]{bls} search for detecting periodic transits produced by smaller planetary companions, but no significant signals were found either.\

We also searched for additional signals in the radial velocities. We ran a GLS periodogram on the residuals of the radial velocities for each system, finding no significant peaks in the corresponding power spectra. Additionally, we included a radial velocity linear trend in the global modelling of each system, but the Bayesian evidences obtained for these models were lower than those of the models not including the linear trend in the radial velocities.

Finally we explored the possibility of identifying long period companions from GAIA astrometric excess noise, but the three systems presented in this study have proper motion uncertainties consistent with other stars having similar magnitudes and parallaxes.

\subsection{Heavy element content}
 \begin{figure}[tp]
    \centering
    \includegraphics[width=\columnwidth]{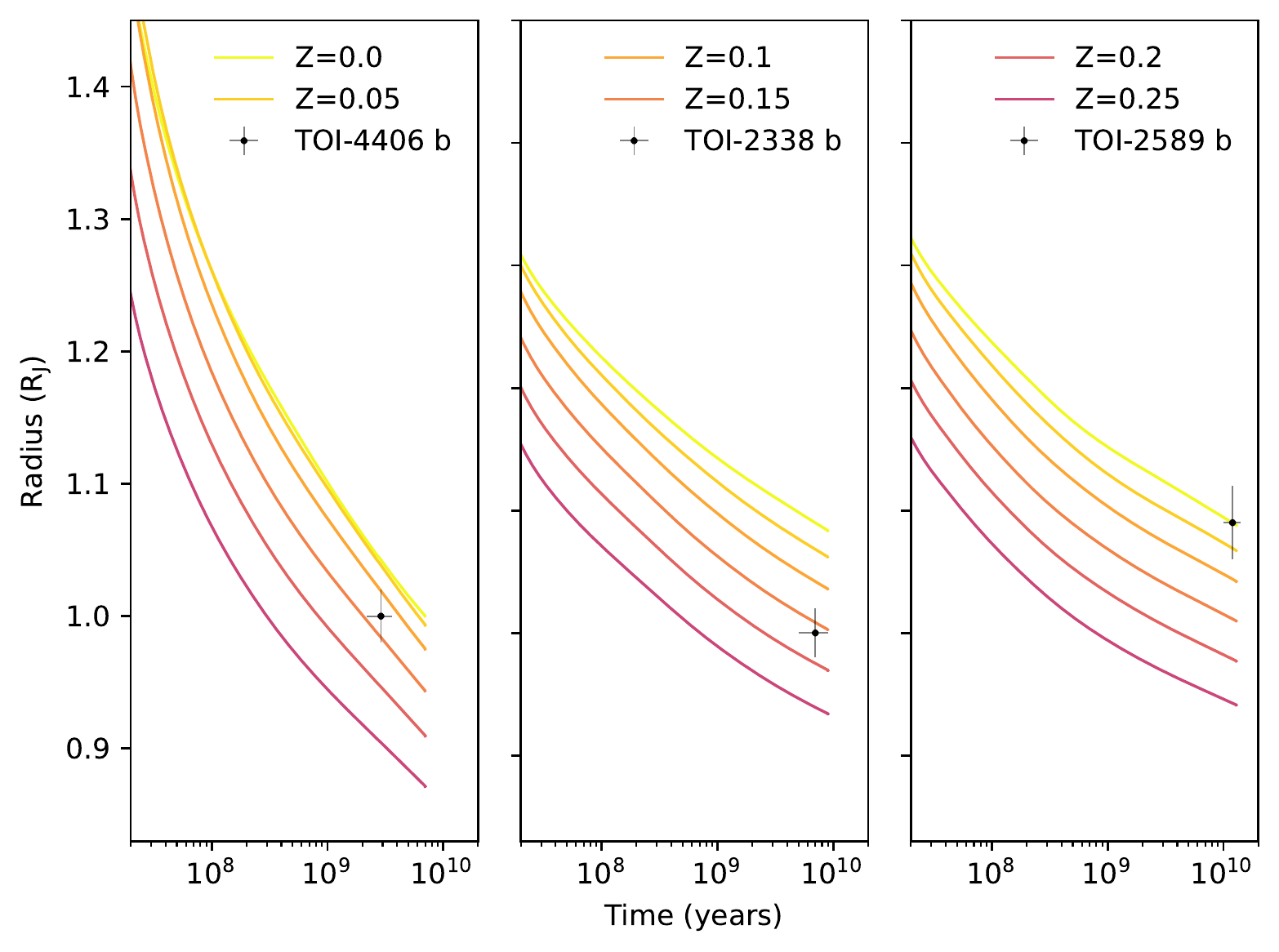}
    \caption{Evolution curves of the planetary radius as a function of time color-coded by the water mass fraction of the envelope. The points correspond to the current properties of each planet}
    \label{heavy_element} 
\end{figure}

We estimate the amount of heavy elements for the three warm Jupiters following the procedure described in \citet{sarkis:2021}.
We compare the planetary radius obtained from the evolution model \texttt{completo21} \citep{mordasini:2012} with the observed radius, while varying the heavy element content of the envelope. \texttt{completo21} has been validated through the modeling of the formation and evolution of Jupiter in the core accretion framework, and the models are in agreement with results presented in \citet{burrows:1997} and \citet{baraffe:2003}.
\texttt{Completo21} is used to model the core and the envelope and is coupled with a semi-grey atmospheric model \citep{guillot:2010}. We chose the SCvH equation of states of hydrogen and helium and a helium mass fraction of Y=0.27 \citep{saumon:1995}. The models of \plA\ and \plB\ contain a core of 10 $M_{\oplus}$ \citep{thorngren:2016} composed of iron and silicates with an iron mass fraction of 33\%. The relatively large radius of \plC\ is compatible with no heavy element enrichment hence we chose not to include a core for this planet. The heavy elements outside the core are modeled by the AQUA2020 EOS of water \citep{haldemann:2020} and are assumed to be homogeneously mixed in the envelope. All planets are modelled from 10 Myrs up to past their current age determination. Figure \ref{heavy_element} displays the resulting evolution curves of the planetary radius for several water mass fractions in the envelope (Z). We derive error estimates on the water mass fraction by a simple Monte Carlo approach combining the planetary radius and stellar age uncertainties, as done in \citet{ulmer:2022}. The parameters of \plA, \plB, and \plC\ are well explained by a total mass of heavy elements of $\rm 21^{+5}_{-5}\,M_{\oplus}$, $\rm 330^{+60}_{-60}\,M_{\oplus}$ and $\rm 40^{+50}_{-40}\,M_{\oplus}$ which corresponds to a total heavy element mass fraction of ($\rm Z_{p}$) of $0.23 ^{+0.05} _{-0.05}$, $0.174 ^{+0.031} _{-0.031}$, and $0.03 ^{+0.04} _{-0.03}$, respectively. Assuming that the stellar metallicity scales with the iron abundance \citep{asplund:2009,miller:2011}, the heavy element enrichment of the planets are the following: $12.6 ^{+3.1} _{-3.1}$ for \plA, $7.4 ^{+1.5} _{-1.5}$ for \plB, and $1.9 ^{+2.3} _{-1.9}$ for \plC. \plA\ and \plB\ are significantly metal-enriched planets while \plC\ is consistent with no heavy element enrichment.

\section{Discussion} \label{sec:disc}
\begin{figure*}[tp]
    \centering
    \includegraphics[width=\textwidth]{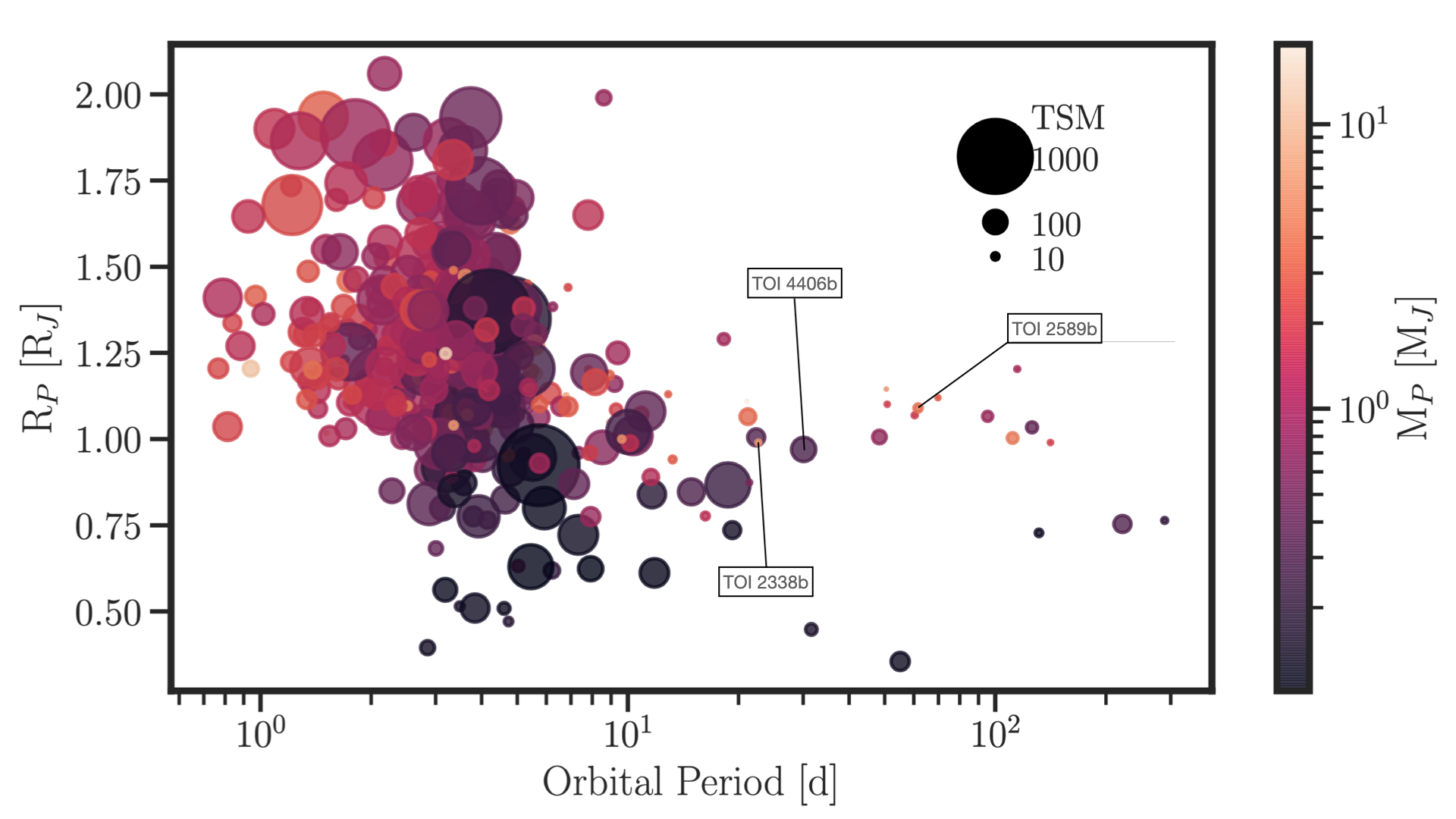}
    \caption{Planet radius versus orbital period diagram for the population of transiting giant planets, color coded by planet mass. The size of the dots scales with the Transit Spectroscopy Metric \citep[TSM,][]{kempton:2018}. The three systems presented in this study stand out due to their relatively long periods.}
    \label{periods} 
\end{figure*}

\plA, \plB, and \plC\ are three transiting Jovian planets with predicted time-averaged equilibrium temperatures below 1000 K, and orbital periods between 22 and 62 days, and can therefore be classified as warm Jupiters. Figure \ref{periods} puts these discoveries in the context of the full population of transiting giant planets, highlighting the relative scarcity of systems with periods longer than 10 days, which is produced mostly by observational biases. \plC, with an orbital period of 61.6 days, is currently among the five longest period confirmed planets identified by the \tess\ mission.

\begin{figure*}
    \centering
    \gridline{\fig{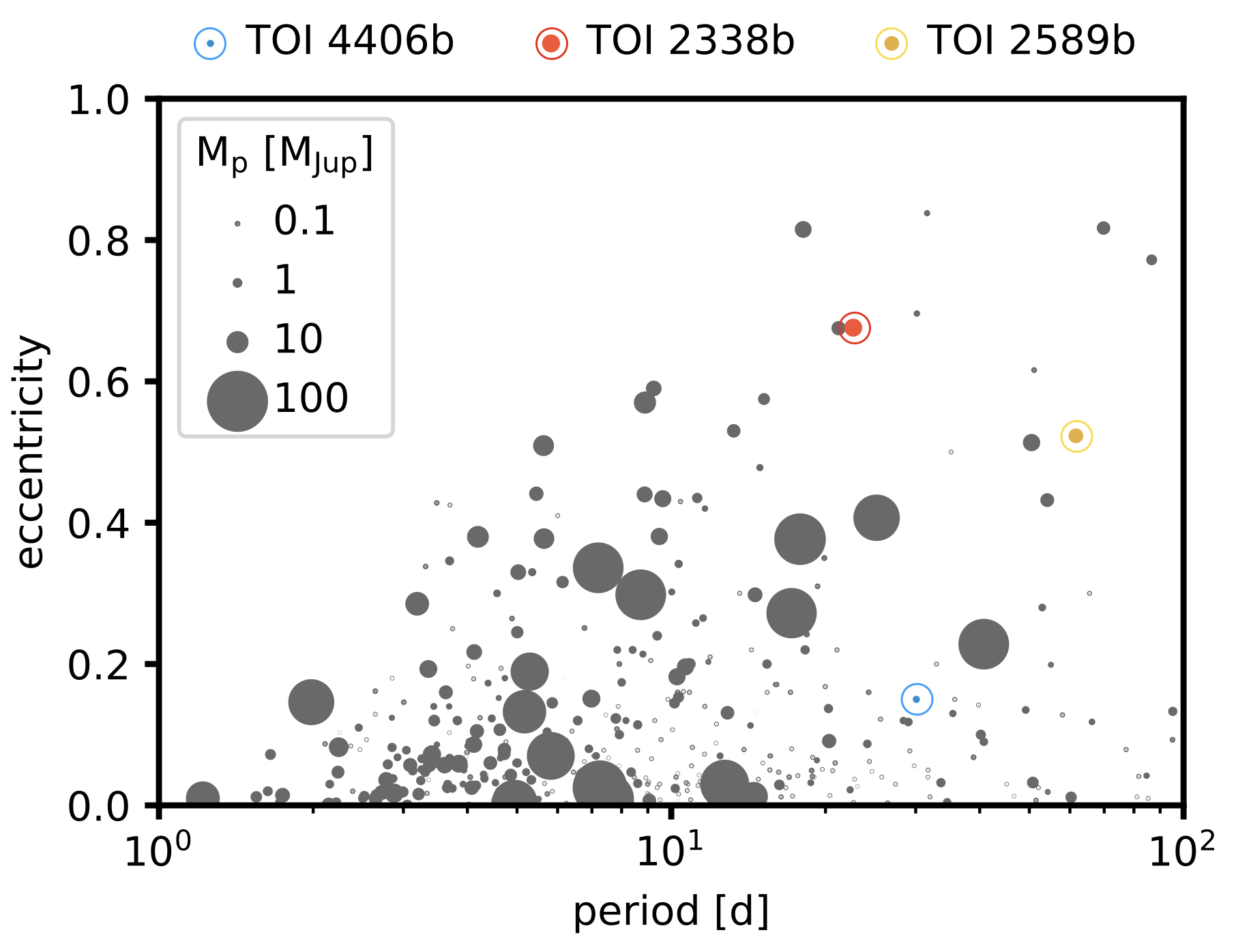}{\columnwidth}{}
              \fig{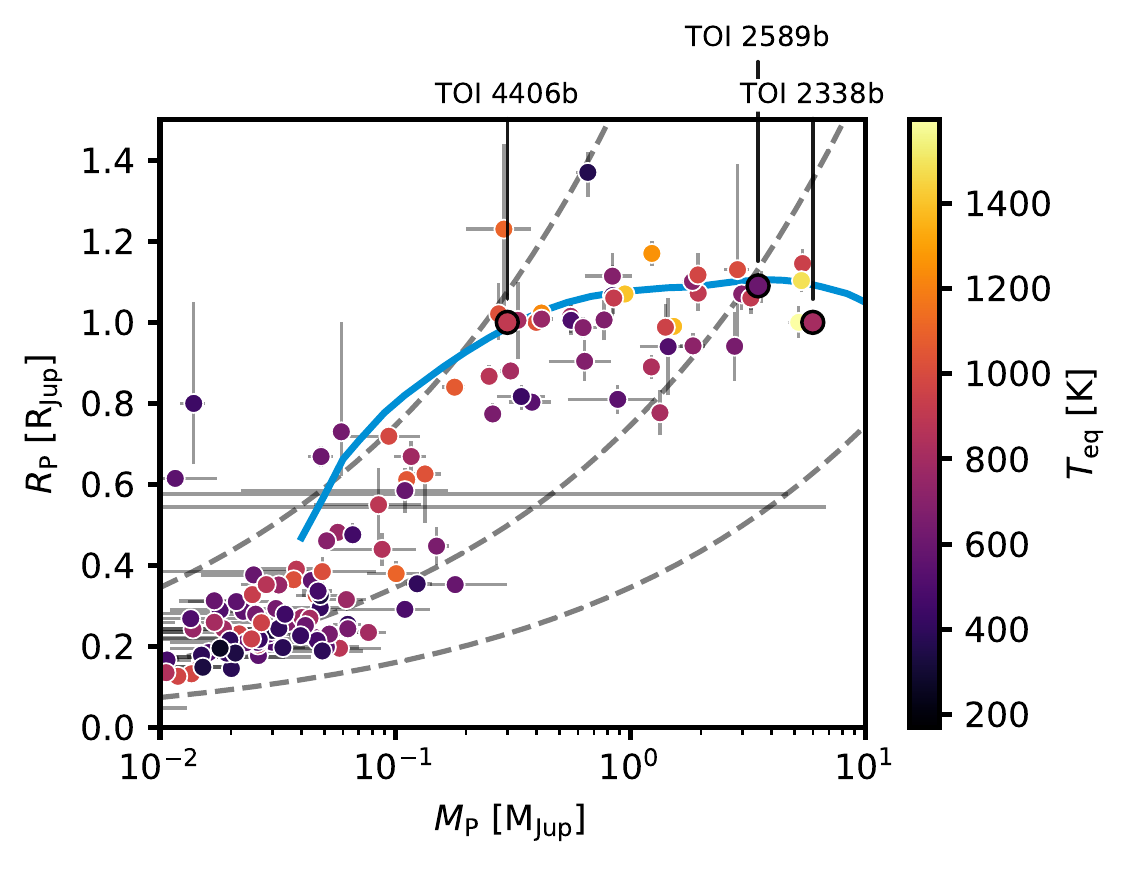}{\columnwidth}{}}
    \caption{\textit{Left panel:} orbital eccentricity as a function of orbital period for the population of transiting giant planets with up to 100~d period. \plB\ and \plC\ are among the few discovered planets with  $e > 0.5$, with \plB\ being in the 98th percentile of the eccentricity distribution. \textit{Right panel:} Planet radius as a function of the planet mass for the population of warm giant planets (P$>$10 days). Dashed gray lines correspond to bulk densities of 0.3, 3, and \SI{30}{\gram\per\centi\meter\cubed}. All three planets presented in this study have radii close to those predicted by standard structural models \citep[][blue line]{fortney:2007}.}
    \label{fig:eccR}
\end{figure*}

Transiting warm Jupiters, due to the mild irradiation levels experienced from their parent stars, allow for the study of their internal bulk structures without the need to consider proximity effects. In our case, the three transiting warm Jupiters have significantly different masses, but similar radii, which is expected given the electron degeneracy pressure for non-inflated envelopes \citep{Zapolsky:1969}. The right panel of Figure \ref{fig:eccR} corresponds to the mass-radius diagram for the population of transiting warm giant planets from the \mbox{TEPCat} catalog~\citep{Southworth2011}. While hot Jupiters have radii well above the predictions based on classical planetary structural models, warm Jupiters have radii similar to or below $\approx$ 1 R$_J$.

The metal enrichment of these three new planets is consistent with the correlation presented in \citet{thorngren:2016}, where lower mass giant planets are more enriched in metals relative to their stars if compared with the more massive giant planets. In this study, we find that the Saturn-mass planet \plA\ is significantly more enriched in metals relative to its star than the two super-Jupiters \plB\ and \plC. While we find that most massive planet of our sample (\plB) has a higher relative enrichment than the less massive super-Jupiter \plC, this deviation is consistent with the moderate scatter presented in the \citet{thorngren:2016} correlation. \plB\ and \plC\ have very similar stellar hosts, and the difference in their internal composition favors the hypothesis that the final structural composition of massive planets is not fully constrained by the properties of the star, but is significantly dependent on specific formation histories. The structure of \plC\ is consistent a core-less interior and an envelope with a stellar-like metal enrichment. These properties can be associated to the gravitational instability formation mechanism, but the mass of \plC\ is below the 10 M$_J$ boundary proposed by \citet{schlaufman:2018} above which objects orbiting solar-type dwarf stars are expected to have been formed by this process. If \plC\ was formed by the core accretion mechanism, then it was able to accrete all the available gas in its feeding zone.

\plB\ and \plC\ stand out due to their large orbital eccentricities (see left panel of Figure \ref{fig:eccR}), while \plA\ has an almost circular orbit. As opposed to hot Jupiters, the population of warm Jupiters is known for having a wide distribution of orbital eccentricities \citep{winn:2015}. The eccentricities can be linked to formation scenarios of close-in giant planets \citep[e.g.][]{dong:2021}. 

\textit{In situ} formation of hot and warm Jupiters presents theoretical challenges:
While growth through planetesimal accretion is fast due to the short orbital timescales, the narrow feeding zones of available material in the inner system are typically depleted before a sufficient core mass can be reached~\citep{Lissauer1993}.
Additional accretion of \SI{}{\milli\meter} to \SI{}{\centi\meter}-sized pebbles~\citep{Ormel2010} can grow a solid core more efficiently~\citep{Batygin2016}, but pebble isolation~\citep[e.g.,][]{Lambrechts2014} limits attainable core masses on close orbits to values too low for runaway gas accretion~\citep{Lin2018}.

An alternative scenario is that these planets form beyond the snow line and then migrate inward through interaction with the gaseous disc~\citep{goldreich:1980,Kley2012}.
Given an appropriate combination of solid disk mass and initial orbital distance of the emerging planetary nucleus, a gas giant on \mbox{sub-au} orbits can form in this way~\citep{schlecker2021b}.

Another mode to form large, close-in planets is excitation of the orbital eccentricities through dynamical interactions followed by orbital shrinking due to tides with the central star~\citep{Goldreich1966}. 
\citet{dong:2021b} recently announced the discovery of a possible hot Jupiter progenitor, a massive warm Jupiter with a highly eccentric orbit (TOI-3362b). While \plB\ has similar orbital and physical parameters to TOI-3362b, its eccentricity is currently not high enough to eventually end up as a hot Jupiter. Assuming a constant angular momentum path \citep[e.g.][]{socrates:2012}, \plB\ is expected to attain a final semi-major axis of $a(1-e^2)\approx 0.08$, but the timescale associated with that process is longer than the lifetime of the star \citep{adams:2006}. This finding is similar to that for TIC~237913194b \citep{schlecker:2020}, which has a similar configuration to \plB\ and whose tidal evolution has been shown to be too slow for a hot Jupiter progenitor.

Under the current orbital configurations, the three planets presented in this study are not expected to become hot Jupiters. Nonetheless, their locations inside the snow line require an explanation, because they need to have experienced significant migration to reach their current semi-major axis.
High eccentricity migration mechanisms triggered by secular gravitational interactions have been proposed as a possible origin for the population of warm Jupiters \citep[e.g.][]{socrates:2012,petrovich:2016,mustill:2017,anderson:2017}. In this scenario, an outer companion periodically modifies the orbital eccentricity and inclination of the inner warm Jupiter, and the migration is only significant in the high eccentricity stages, when the pericenter distances are small enough to allow for strong tidal interactions that reduce the semi-major axes of the orbit. The large eccentricities of \plB\ and \plC\ indicate that they could be experiencing such secular cycles. On the other hand, the small orbital eccentricity of \plA\ is consistent with having experience disc migration until reaching its current position.

\citet{dong:2014} pointed out that in order for warm Jupiters to be migrating through Kozai-Lidov oscilations triggered by an outer jovian mass companion, this  second planet must be located at relatively close orbital distance in order to overcome the the oscillation suppressing effect of general relativity \citep{fabrycky:2007,liu:2015}.
Such companions should be detectable in long term radial velocity monitoring of these systems. \citet{jackson:2021} predicted that such radial velocity trends induced by the outer perturbers should be detectable in a time span of 3 years. Both eccentric systems presented in this study have been monitored for three years but no trend has been detected, ruling out the existence of an important fraction of possible perturbers. An extended and more precise radial velocity monitoring of these systems will be required to firmly reject the existence of outer perturbers that could be responsible for the migration. 

In the absence of such close perturbers, the current orbital configurations of \plB\ and \plC\ can be explained by an initial stage of migration through the disc of systems composed of multiple giant planets, followed by a stage of scattering between the planets after the disc disperses. Based on  N-body experiments, \citet{anderson:2020} show that the majority of eccentric warm Jupiters with no detected companions can be explained by this in-situ scattering process, which supports this avenue for the current orbital state of \plB\ and \plC.
 
A crucial test to constrain the origin of the warm Jupiters presented in this study is through the measurement of their orbital inclinations. Specifically, migration through secular mechanisms predicts significant inclinations between the orbital plane of the warm Jupiter and the spin of the star (stellar obliquity).
 The sky projected stellar obliquity can be obtained with radial velocity measurements during transit through the observation of the Rossiter-McLaughlin (R-M) effect \citep[e.g.][]{Sedaghati:2021,dong:2022}. The expected maximum radial velocity amplitude of the R-M effect is similar for the three systems. We predict amplitudes of the R-M signals for \plA\ and \plC\ of $\approx$9 m/s, for aligned orbits. For \plB, due to the smaller impact parameter of its transit, we predict a slightly larger signal of 16 m/s in the aligned case. The transit duration in all three  cases (shorter than 7 hours) should allow for the spectroscopic observation of a full transit in a single night. A significant obliquity value for any of the three systems would further favor the high eccentricity migration scenario triggered by secular gravitational interactions with outer companions as the cause of the close-in orbits \citep{petrovich:2016}. \citet{rice:2022b} found a tendency toward spin-orbit alignment of warm-Jupiter systems, and suggest it as evidence of distinct migration mechanisms for the hot and warm Jupiter populations. The obliquity measurement of a bigger sample of warm Jupiters, particularly those that are eccentric like \plB\ and \plC, will help in confirming this claim.

Formation and migration scenarios of giant planets can also be constrained by studying their atmospheric compositions~\citep[e.g.,][]{Oberg2011,Mordasini2016,Molliere2022}. For example, \citet{Madhusudhan:2014} inferred that high-eccentricity migration after disk dissipation can lead to sub-stellar carbon and oxygen abundances with stellar or superstellar C/O abundance ratios. Planetary atmospheric abundances can be constrained with transmission spectroscopy measurements and also with the observation of the emission spectrum during secondary eclipses. The compact nature of \plB\ and \plC\ does not favour atmospheric studies in transmission, but emission studies would be possible. \plA, on the other hand, is a well suited target for both transmission and emission studies due to its lower density. As can be noted in Figure \ref{periods}, \plA\ has currently the highest transmission spectroscopy metric \citep{kempton:2018} for giant planets with periods longer than 20 days.

\acknowledgments
R.B. acknowledges support from FONDECYT Project 11200751. A.J., R.B., F.R., P.T., and M.H.\ acknowledge support from ANID -- Millennium  Science  Initiative -- ICN12\_009. A.J.\ acknowledges additional support from FONDECYT project 1210718.
T.T. acknowledges support by the DFG Research Unit FOR 2544 "Blue Planets around Red Stars" project No. KU 3625/2-1.
T.T. further acknowledges support by the BNSF program "VIHREN-2021" project No. КП-06-ДВ/5.
The results reported herein benefited from collaborations and/or information exchange within the program “Alien Earths” (supported by the National Aeronautics and Space Administration under agreement No. 80NSSC21K0593) for NASA’s Nexus for Exoplanet System Science (NExSS) research coordination network sponsored by NASA’s Science Mission Directorate.
The contributions of SU, FB, AD, GC, NG and ML have been carried out within the framework of the NCCR PlanetS supported by the Swiss National Science Foundation under grants 51NF40\_182901 and 51NF40\_205606. The authors acknowledge the financial support of the SNSF. ML acknowledges support of the Swiss National Science Foundation under grant number PCEFP2\_194576.
This research received funding from the European Research Council (ERC) under the European Union's Horizon 2020 research and innovation programme (grant agreement n$^\circ$ 803193/BEBOP), and from the Science and Technology Facilities Council (STFC; grant n$^\circ$ ST/S00193X/1).
We acknowledge the use of public TESS data from pipelines at the TESS Science Office and at the TESS Science Processing Operations Center. Resources supporting this work were provided by the NASA High-End Computing (HEC) Program through the NASA Advanced Supercomputing (NAS) Division at Ames Research Center for the production of the SPOC data products.
This publication was made possible through the support of an LSSTC Catalyst Fellowship to T.D., funded through Grant 62192 from the John Templeton Foundation to LSST Corporation. The opinions expressed in this publication are those of the authors and do not necessarily reflect the views of LSSTC or the John Templeton Foundation.
The postdoctoral fellowship of KB is funded by F.R.S.-FNRS grant T.0109.20 and by the Francqui Foundation.
The ASTEP team acknowledges support from IPEV, PNRA, the Université Côte d’Azur, the University of Birmingham and ESA. 
This work makes use of observations from the LCOGT network. Part of the LCOGT telescope time was granted by NOIRLab through the Mid-Scale Innovations Program (MSIP). MSIP is funded by NSF.
Funding for the \tess\ mission is provided by NASA’s Science Mission directorate. We acknowledge the use of public \tess\ Alert data from pipelines at the \tess\ Science Office and at the \tess\ Science Processing Operations Center.

\facilities{TESS, GAIA, LCO, El Sauce, ASTEP, Hazelwood Observatory, SOAR, ESO 3.6m, MPG 2.2m, SMARTS 1.5m, Euler-Swiss telescope 1.2 m.}

\software{astropy \citep{2013A&A...558A..33A},  
          \texttt{ceres} \citep{brahm:2017:ceres},
          \texttt{AstroImageJ}
          \citep{collins:2017},
          \texttt{zaspe} \citep{brahm:2016:zaspe},
          \texttt{juliet} \citep{espinoza:2019},
          \texttt{batman} \citep{kreidberg:2015},
          \texttt{radvel} \citep{fulton:2018},
          \texttt{emcee} \citep{emcee:2013},
          \texttt{dynesty} \citep{dynesty}.
          }
\newpage
\bibliography{sample63}{}
\bibliographystyle{aasjournal}

\end{document}